\documentclass[aps, pre, showpacs, onecolumn]{revtex4-1} 
\usepackage{amsmath}
\usepackage{amssymb} 
\usepackage{graphicx} 
\usepackage[caption=false]{subfig}
\usepackage{epstopdf}
\usepackage{natbib} 
\begin{document}

\begin{abstract} We introduce and study a model of time-dependent billiard
systems with billiard boundaries undergoing infinitesimal wiggling motions. The
so-called quivering billiard is simple to simulate, straightforward to analyze,
and is a faithful representation of time-dependent billiards in the limit of
small boundary displacements. We assert that when a billiard's wall motion
approaches the quivering motion, deterministic particle dynamics become
inherently stochastic. Particle ensembles in a quivering billiard are shown to
evolve to a universal energy distribution through an energy diffusion process,
regardless of the billiard's shape or dimensionality, and as a consequence
universally display Fermi acceleration. Our model resolves a known discrepancy
between the one-dimensional Fermi-Ulam model and the simplified static wall
approximation. We argue that the quivering limit is the true fixed wall 
limit of the Fermi-Ulam model.
\end{abstract}

\title{Universal energy diffusion in a quivering billiard} 
\author{Jeffery Demers}
\author{Christopher Jarzynski}
\affiliation{University of Maryland, College Park, Maryland, 20742, USA}
\date{14 September 2015}
\pacs{05.45.-a, 05.40.-a}
\maketitle

\section{Introduction}
Billiards are remarkably useful physical models; they allow a diverse range of 
classical dynamics to be understood intuitively through easy-to-visualize 
particle trajectories and are a natural setting for quantum and wave chaos
\cite{Ott2002}, while the discrete time nature of particle-billiard boundary
interactions make classical billiards especially amenable to numerical study.
Time-dependent billiards (billiards with boundaries in motion) in particular 
can be found in a wide range of applications: KAM theory \cite{Brahic, L&L1972,
LLC1980}, one-body dissipation in nuclear dynamics \cite{JarSwia1993}, Fermi
acceleration \cite{Fermi1949, Z&C, Ulam, Hammersley, L&L1972, Brahic, GRT2012,
BAT2014A, BAT2014B}, and adiabatic energy diffusion \cite{Wilkinson1990,
Jar1993}, for example.
\\ 
\indent This work was originally motivated by the
desire to study and simulate classical particle trajectories in time-dependent
billiard systems. The task is complicated by the boundary's displacement, which
produces implicit equations for the time between particle-boundary collisions.
We propose a fixed wall simplification by considering the limit of
infinitesimally small boundary displacements. Our limit will be called the
quivering limit, and the resulting billiard system will be called a quivering
billiard. The purpose of this paper is to show that, although simple, quivering
billiards are accurate descriptions of time-dependent billiards in the limit of
small boundary displacements, and to support our conjecture that any
physically consistent, non-trivial, fixed wall simplification of a
time-dependent billiard must be physically equivalent to a quivering billiard. 
Using physical reasoning, we will argue that in the
quivering limit, deterministic billiard dynamics become inherently stochastic.
Then, by utilizing the simplifications allowed by stochastic methods and fixed
billiard walls, we will derive analytic expressions to describe energy evolution
in a quivering billiard. Our investigations will uncover universal behavior in
time-dependent billiards when billiard motion is close to the quivering limit,
and our results will enable us to addresses several issues that have been raised
in previous Fermi acceleration and time-dependent billiard literature. \\
\indent The outline of this paper is as follows. In Sec.~\ref{sec:II}, we first
define a quivering billiard and determine its behavior in one dimension, and
then generalize to quivering billiards in arbitrary dimensions. The energy
statistics of a single particle and a particle ensemble are examined in
Sec.~\ref{sec:III}, and the results are discussed in the context previous
literature in Sec.~\ref{sec:IV}. In Sec.~\ref{sec:V}, we give examples of
quivering billiards and present numerical analyses, and we conclude in
Sec.~\ref{sec:VI}.

\section{The Quivering Limit} \label{sec:II} In this section, we define
quivering as a particular limit of time-dependent billiard motion. Because the
dynamics are so poorly behaved in this limit, billiard systems can only be
described stochastically. For simplicity, we first work with a one-dimensional
billiard with a single moving wall, and then extend to arbitrary billiard motion
in arbitrary dimensions.

\subsection{The 1-D Fermi-Ulam Model} \label{sec:IIA} \indent Consider a
particle in one dimension bouncing between two infinitely massive walls. One
wall is fixed at $x=0$, and the other oscillates about its mean position at $x =
L$, where we take $L>0$. The particle's energy fluctuates due to collisions with
the moving wall, and the dynamical system corresponding to the particle's motion
defines the well-known Fermi-Ulam model 
\cite{Ulam,Hammersley,Z&C,Brahic,L&L1972}. Suppose that the moving wall
oscillates periodically with period $\tau$, characteristic oscillation amplitude
$a$, and characteristic speed $u_{c} = a / \tau$. The moving wall's position
$x(t)$ and velocity $u(t)$ at time $t$ can be written as 
\begin{eqnarray}\label{WallEqs1} 
x(t) & = & L + g(t) \\ 
u(t) & = & \frac{\mathrm{d}g}{\mathrm{d}t}, \nonumber 
\end{eqnarray} 
where $g(t)$ is some piecewise smooth $\tau$-periodic function with mean zero. 
The wall velocity scales like $u_{c}$, and $g(t)$ scales like $a$. To make the 
scaling obvious, we note that $g(t)$ depends on $t$ only through the value of 
$t$ mod $\tau$, and we make the following substitutions: 
\begin{eqnarray}\label{defs1}
\Psi(t) & = &\frac{t}{\tau}~\mathrm{mod}~1 \\ 
g(t) & = & a \, h \boldsymbol{(} \Psi(t)\boldsymbol{)}. \nonumber 
\end{eqnarray} 
The quantity $\Psi(t)$ will be referred to as the wall's phase. Here, $h$ is 
regarded as a function of $\Psi$, and $h\boldsymbol{(}\Psi(t)\boldsymbol{)}$ 
means $h(\Psi)$ evaluated for $\Psi =\Psi(t)$. The quantity 
$h\boldsymbol{(}\Psi(t)\boldsymbol{)}$ is just $g(t)$ rescaled to have a 
characteristic oscillation amplitude of unity. The state of the wall at time 
$t$ is thus 
\begin{eqnarray} \label{WallEqs2}
x(t) & = & L + a~h\boldsymbol{(}\Psi(t)\boldsymbol{)} \\
u(t) & = & u_{c}\,h'\boldsymbol{(}\Psi(t)\boldsymbol{)}, \nonumber 
\end{eqnarray} 
where the $h'$ denotes the derivative of $h$ with respect to its argument 
$\Psi$. 
\\
\indent We define the quivering limit of the Fermi-Ulam model by taking $a, \tau
\rightarrow 0$ while holding $u_{c}$ constant and leaving the dependence of $h$
on $\Psi$ fixed. In the quivering limit, the moving wall's position reduces to
$x(t) = L$, so no implicit equations for the time between collisions arise from
the dynamics. This simplification comes at a price; when $\tau \rightarrow 0$,
$\Psi$ oscillates infinitely fast in time, and $u(t)$ does not converge to any
value for any given $t$. That is, in the quivering limit, $u(t)$ becomes
ambiguous to evaluate. Our task now is to physically interpret and resolve this
ambiguity. 
\\ 
\indent Note that in the quivering limit, the wall makes
infinitely erratic motions at finite speeds; the $n^{th}$ derivative of $g(t)$,
scaling like $a / \tau^{n}$, diverges for all $n \geq 2$. An infinitesimal
change in the state of a particle results in a finite and essentially
unpredictable change in the wall's velocity at the time of the next bounce. We
assert that one could never, even in principle, specify the state of the
particle with enough precision to reliably predict the velocity of the moving
wall, and thus the change in particle energy, during the next collision. We
therefore claim that in the quivering limit, the dynamics of the Fermi-Ulam
model become inherently stochastic; deterministic particle trajectories defined
on phase space transition to stochastic processes defined on a probability
space. Given any initial condition, the resulting particle trajectory actually
represents one possible realization drawn from an ensemble of initial conditions
infinitesimally displaced from one another. The wall's velocity during a
collision will be treated as a random variable, and we now find the
corresponding probability distribution. 
\\ 
\indent Consider again the moving
wall with non-zero $a$ and $\tau$. Let $P(u|0)$ be the probability density for a
stationary observer to measure the velocity $u$ during a randomly timed snapshot
of the wall: 
\begin{eqnarray} \label{P(u|0)} 
P(u|0) & = & \frac{1}{\tau}\int_0^{\tau} \mathrm{d}t\,\delta\boldsymbol{(}u -
u(t)\boldsymbol{)} \\
 & = & \int_{0}^{1} \mathrm{d}\Psi \delta\boldsymbol{(}u - 
 u_{c}\,h'(\Psi)\boldsymbol{)}. \nonumber 
\end{eqnarray}
The reason for placing the conditional 
$|0$ in the argument of $P$ will become apparent shortly. We note that 
$P(u|0)$ is normalized, so it is indeed a well-defined probability 
density. In the quivering limit, $u_{c}$ and the dependence of $h$ on $\Psi$ 
remain constant, so $P(u|0)$ remains well-defined and unchanged. If 
the stationary observer were to measure the wall velocity in the quivering 
limit, any observation, no matter how well-timed, would be an essentially 
random snapshot due to the wall's infinitely erratic motion. We thus take 
$P(u|0)$ to be the probability for a stationary observer to measure the wall 
with velocity $u$ when the wall is quivering. 
\\
\begin{figure} 
\includegraphics{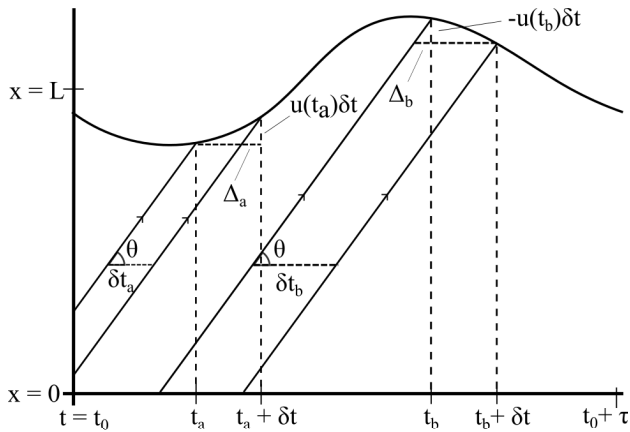}
\caption{\label{Spacetime diagram}Spacetime diagram over one period of a moving
wall's motion. The smooth curve represents the wall's position, and particles
approach the wall along the diagonal arrows to collide at times $t_{a}, t_{a} +
\delta t, t_{b}$ and $t_{b} + \delta t$} 
\end{figure} 
\indent The particle bouncing between the walls effectively measures the wall's 
velocity during collisions, but the particle is not a stationary observer. 
Collisions with large relative speeds of approach occur more frequently than 
collisions with small relative speeds of approach, so there exists a 
statistical bias that favors collisions for which the wall moves towards the 
particle. If the quivering dynamics are to be physically consistent with the 
Fermi-Ulam dynamics, this statistical bias must be incorporated into the 
probability distribution used to determine the wall's velocity during 
collisions. The mathematical realization of the statistical bias can be found 
with the aid of Fig.~\ref{Spacetime diagram}, a construction first employed by 
Hammersley~\cite{Hammersley} and Brahic~\cite{Brahic}. 
\\ 
\indent In Fig.~\ref{Spacetime diagram}, the position of the moving wall in the 
Fermi-Ulam model is plotted over one period of motion in the interval 
$(t_{0}, t_{0} + \tau)$. Consider an ensemble of particles approaching the 
moving wall with speed $v$. For the moment, we assume that $v$ is larger than 
the maximum wall velocity $u_{max}$. The particles are launched from $x = 0$ at 
a uniform rate over a period of duration $\tau$ such that they all collide 
with the wall during the interval $(t_{0}, t_{0} + \tau)$. We concern 
ourselves only with the first collision each particle makes with the
moving wall. Four trajectories from the ensemble are shown in
Fig.~\ref{Spacetime diagram}, representing collisions with the wall at times
$t_{a}, \,t_{a} + \delta t, \, t_{b},$ and $t_{b} + \delta t$. Because the
launch times are uniformly distributed, the fraction of particles that collide
with the wall between $t_{a}$ and $t_{a} + \delta t$ will be proportional to the
interval $\delta t_{a} = \delta t - \Delta_{a}$. Likewise, the fraction that
collide between $t_{b}$ and $t_{b} + \delta t$ will be proportional to $\delta
t_{b} = \delta t + \Delta_{b}$. Using the geometry of Fig.~\ref{Spacetime
diagram} and the fact that $\tan(\theta) = v$, we find the probability density
for randomly selected ensemble member collide with the moving wall at a time $t$
within the interval $(t_{0}, t_{0} + \tau)$ to be
\begin{equation} \label{P_u(t)1d} 
P\boldsymbol{(}u(t) | v\boldsymbol{)} =
\frac{1}{\tau}\left(1-\frac{u(t)}{v}\right). 
\end{equation}
Multiplying by a delta function and integrating Eq.~(\ref{P_u(t)1d}) over a 
period of the wall's motion gives $P(u|v)$, the probability density for a 
randomly selected ensemble member's collision to occur when the wall 
moves with velocity $u$:
\begin{eqnarray} \label{P_v(u) 1d half} 
P(u|v) & = & \frac{1}{\tau} \int\limits_{0}^{\tau}\mathrm{d}t\,
\delta\boldsymbol{(}u -u(t)\boldsymbol{)}\left(1-\frac{u(t)}{v}\right) \\ 
& = & \int\limits_{0}^{1}\mathrm{d}\Psi \, \delta\boldsymbol{(}u -
u_{c}\,h'(\Psi)\boldsymbol{)}\left(1-\frac{u_{c}\,h'(\Psi)}{v}\right) 
\nonumber \\ 
& = & P(u|0)\left( 1 - \frac{u}{v}\right). \nonumber 
\end{eqnarray}
Because the wall's average displacement over one period of motion is zero, the 
product $u P(u|0)$ integrated over all wall velocities must also give zero, 
and $P(u|v)$ is therefore normalized and a well-defined probability density. 
The distribution $P(u|v)$ has a statistical bias towards larger negative $u$ 
due to the flux factor $1-u/v$. We will henceforth refer to $P(u|0)$ as the 
unbiased distribution and $P(u|v)$ as the biased distribution. In the 
quivering limit, $P(u|v)$ remains well-defined and unchanged. As 
$\tau \rightarrow 0$, an ensemble of particles launched over a period of wall 
motion from a fixed $x$ is essentially equivalent to an ensemble of 
infinitesimally displaced initial conditions. We therefore take $P(u|v)$ to be 
the conditional probability density to observe a quivering wall with velocity 
$u$ during a collision, given that the particle approaches the wall with speed 
$v > u_{max}$.
\\ 
\indent If a particle approaches the moving wall with speed $v < u_{max}$, 
then $P(u|v)$ will become negative for some values of $u$, and 
Eq.~(\ref{P_v(u) 1d half}) will make no sense as a probability density. These 
$u$ values correspond to impossible collisions for which the wall moves with 
positive velocity away from the particle faster than the particle moves toward 
the wall. Such collisions occur with probability zero, and we can account 
for this by simply attaching a step-function to the biased distribution, yielding 
\begin{equation}
\label{P_v(u) 1d} P(u|v) = 
\begin{cases} 
P(u|0)\left(1-\frac{u}{v}\right), & v \geq u_{max} \\ 
N(v)P(u|0)\left(1-\frac{u}{v}\right)\Theta(v-u), & v < u_{max},
\end{cases} 
\end{equation} 
where $\Theta(x)$ is the unit step function (equal to
$0$ for $x<0$ and $1$ for $x \geq 0$) and $N(v)$ is a $v$ dependent
normalization. 
\\ 
\indent Equation (\ref{P_v(u) 1d}) determines the statistics
of a particle's energy evolution in a quivering Fermi-Ulam system. As with any
billiard system, the particle's energy is simply the kinetic energy
$\frac{1}{2}m v^{2}$, where $m$ is the particle's mass and $v$ is its speed. The
particle bounces between the two walls as if the system were time-independent,
but when colliding with the quivering wall at an incoming speed $v_{i}$ (the
particle moves in the positive $x$ direction to collide with the moving wall, so
$v_{i}$ is also the incoming velocity), a value for the wall velocity $u$ is
selected using the biased distribution $P(u|v_{i})$. The particle's velocity
just after the collision, $v_{f}$, is given by 
\begin{equation}\label{v_f_1d}
v_{f} = 2u-v_{i}, 
\end{equation} 
and the corresponding energy change, $\Delta E$, is given by 
\begin{equation} \label{delta_E_1d} 
\Delta E =2mu^{2}-2mu\,v_{i}. 
\end{equation} 
Equations (\ref{v_f_1d}) and (\ref{delta_E_1d}) are determined using the 
standard collision kinematics for a particle in one-dimension colliding 
elastically with an infinitely massive moving object.
\\ 
\indent Before moving on to higher dimensions, we must address
the possibility of particles escaping the billiard interior. This issue will
plague any fixed wall simplification of time-dependent billiards, and is
discussed in detail in Ref.~\cite{KDC2008}. From Eq.~(\ref{v_f_1d}), we see that
if $0 < u < v_{i} \leq 2u$, the particle does not turn around after a collision with
the moving wall, but instead slows down and continues forward. We refer to these
types of collisions as glancing collisions. For non-zero $a$ and $\tau$, just
after a glancing collision, the particle continues forward slower than the wall
moves outward, so the particle will remain within the billiard interior. With a
fixed wall simplification, however, the wall does not actually move outward
after a glancing collision, so the particle will continue forward and escape the
billiard interior. A particle escaping through a hard wall is a non-physical
by-product of setting $a=0$, so in order to make a physically reasonable fixed
wall simplification, one must always devise a method to handle glancing
collisions. Our method for a quivering Fermi-Ulam system is devised as follows.
\\ 
\indent For non-zero $a$ and $\tau$, after a glancing collision occurs, the
wall continues to evolve through its period, and one of two possibilities will
occur. The wall may slow down sometime after the glancing collision and allow
the particle to catch up and collide again, or the wall may reverse its
direction and move inward sometime after the glancing collision, also allowing
the particle to collide again. In either case, a second collision occurs after
the first collision, and as $a$ and $\tau$ approach zero, the second occurs
essentially instantaneously after the first. Therefore, we treat a glancing
collision in a quivering Fermi-Ulam billiard as a double collision. When a
particle with speed $v_{i}$ (also the particle's velocity) collides with the
quivering wall, we draw a $u$ value from the distribution $P(u|v_{i})$. If the
selected value of $u$ is such that $0<u<v_{i}\leq 2u$, the particle's new speed
$v_{f}$ (also velocity) is given by $v_{f} = 2u -v_{i}$, and we draw a new $u$
value from the distribution $P(u|v_{f})$. If the second $u$ value gives another
glancing collision, we again update the particle's speed and then draw a third
$u$ value. The process is repeated until a non-glancing collision occurs, and
the whole event (which occurs instantaneously) is treated as a single
collision.

\subsection{Arbitrary Time-Dependent Billiards} \label{sec:IIB} 
We now generalize to arbitrary billiards in arbitrary dimensions. Consider a
time-dependent billiard in $d$ dimensions moving periodically through some
continuous sequence of shapes with period $\tau$, characteristic oscillation
amplitude $a$, and characteristic speed $u_{c} = a/ \tau$. The evolution of any
one point on the boundary will be denoted by the path $\mathbf{q}(t)$, where
$\mathbf{q}(t + \tau) = \mathbf{q}(t)$. For every $t$, the set of all boundary
points $\{\mathbf{q}(t)\}$ is assumed to define a collection of unbroken $d-1$ 
dimensional
surfaces, which we refer to as the boundary components, enclosing some $d$ 
dimensional bounded connected volume. The outward 
unit normal to the billiard boundary at the 
point $\mathbf{q}(t)$ is denoted by 
$\mathbf{\hat{n}}\boldsymbol{(}\mathbf{q}(t)\boldsymbol{)}$, and the
velocity of the boundary point $\mathbf{q}(t)$ is denoted by
$\mathbf{u}\boldsymbol{(}\mathbf{q}(t)\boldsymbol{)} = \mathrm{d}\mathbf{q}(t)
/\mathrm{d}t$. The billiard shape evolves continuously in time, and we assume
that the boundary components remain unbroken throughout their evolution, so
$\mathbf{u}\boldsymbol{(}\mathbf{q}(t)\boldsymbol{)}$ forms a smooth  
vector
field with domain on the boundary $\{\mathbf{q}(t)\}$ for any fixed $t$. 
Likewise, $\mathbf{\hat{n}}\boldsymbol{(}\mathbf{q}(t)\boldsymbol{)}$ forms a 
smooth field on $\{\mathbf{q}(t)\}$ for any fixed $t$, except possibly 
at corners, where $\mathbf{\hat{n}}\boldsymbol{(}\mathbf{q}(t)\boldsymbol{)}$ 
is ill-defined and discontinuous.  We
denote the outward normal velocity of the point $\mathbf{q}(t)$ by
$u\boldsymbol{(}\mathbf{q}(t)\boldsymbol{)} =
\mathbf{u}\boldsymbol{(}\mathbf{q}(t)\boldsymbol{)} \cdot
\mathbf{\hat{n}}\boldsymbol{(}\mathbf{q}(t)\boldsymbol{)}$. 
\\ 
\indent Denote by $\mathbf{q}$ the average of $\mathbf{q}(t)$ over one period: 
\begin{equation}
\mathbf{q} = \frac{1}{\tau}\int_{0}^{\tau} \mathrm{d}{t} \, \mathbf{q}(t).
\end{equation} 
Noting that the boundary components remain unbroken throughout the period of 
motion, it is straightforward to show that set of average boundary points
$\{\mathbf{q}\}$ forms a collection of unbroken $d-1$ dimensional surfaces. 
The trajectory $\mathbf{q}(t)$ and normal velocity
$u\boldsymbol{(}\mathbf{q}(t)\boldsymbol{)}$ of any given boundary point can be
written as functions of the corresponding average location $\mathbf{q}$ and the
time $t$: 
\begin{eqnarray} \label{WallEqs3} 
\mathbf{q}(t) & = & \mathbf{q} + \mathbf{g}(\mathbf{q}, t) \\ 
u(\mathbf{q},t) & = & \partial_{t}\mathbf{g}(\mathbf{q}, t) \cdot
\mathbf{\hat{n}}\boldsymbol{(}\mathbf{q}(t)\boldsymbol{)}, \nonumber
\end{eqnarray} 
where $\mathbf{g}(\mathbf{q},t)$ is a piecewise smooth in time
$\tau$ periodic function with a time average of zero. $\mathbf{g}(\mathbf{q},t)$
scales like $a$ and $u\boldsymbol{(}\mathbf{q}(t)\boldsymbol{)}$ scales like
$u_{c}$. Equation (\ref{WallEqs3}) depends on $t$ only through the value of
$\Psi(t) = t / \tau~\mathrm{mod}~1$, so we write 
\begin{eqnarray}\label{WallEqs4} 
\mathbf{q}(t) & = & \mathbf{q} + a \, \mathbf{h}\boldsymbol{(}\mathbf{q}, 
\Psi(t)\boldsymbol{)} \\
u(\mathbf{q},t) & = & u_{c} \, \partial_{\Psi}\mathbf{h}\boldsymbol{(}\mathbf{q}
,\Psi(t)\boldsymbol{)} \cdot \mathbf{\hat{n}}\boldsymbol{(}\mathbf{q}(t)
\boldsymbol{)}. \nonumber
\end{eqnarray} 
where $a \, \mathbf{h}\boldsymbol{(}\mathbf{q},
\Psi(t)\boldsymbol{)} = \mathbf{g}(\mathbf{q},t)$. Analogously to the one
dimensional case, $\mathbf{h}$ is regarded as a function of $\mathbf{q}$ and
$\Psi$, and $\mathbf{h}\boldsymbol{(}\mathbf{q}, \Psi(t)\boldsymbol{)}$ means
$\mathbf{h}(\mathbf{q},\Psi)$ evaluated for $\Psi = \Psi(t)$. The quivering
limit of an arbitrary dimensional billiard is defined by taking $a, \tau
\rightarrow 0$ while holding $u_{c}$ and the dependence of $\mathbf{h}$ on
$\Psi$ and $\mathbf{q}$ constant. In this limit, the billiard's boundary points
become fixed in time at the average locations $\{\mathbf{q}\}$, so the outward
normal vectors become fixed in time as well. Thus, in the quivering limit, we
have 
\begin{eqnarray} \label{WallEqs5} 
\mathbf{q}(t) & = & \mathbf{q} \\
u(\mathbf{q},t) & = & u_{c}
\,\partial_{\Psi}\mathbf{h}\boldsymbol{(}\mathbf{q},\Psi(t)\boldsymbol{)} \cdot
\mathbf{\hat{n}}(\mathbf{q}) \nonumber \\ 
& = & u_{c}\,h'\boldsymbol{(}\mathbf{q}, \Psi(t)\boldsymbol{)}, \nonumber
\end{eqnarray} 
where we write $h'\boldsymbol{(}\mathbf{q}, \Psi(t)\boldsymbol{)}
= \partial_{\Psi}\mathbf{h}\boldsymbol{(}\mathbf{q}, \Psi(t)\boldsymbol{)} \cdot
\mathbf{\hat{n}}(\mathbf{q})$ for brevity. Any time-dependent billiard taken to
the quivering limit will be called a quivering billiard. 
\\ 
\indent Analogously to the one dimensional case, we define the unbiased 
distribution for each $\mathbf{q}$: 
\begin{equation} \label{P(u|0)nd} 
P(u|0, \mathbf{q}) = \int_{0}^{1} \mathrm{d}\Psi \delta\boldsymbol{(}u -
u_{c}\,h'(\mathbf{q},\Psi)\boldsymbol{)}. 
\end{equation} 
\begin{figure}
\includegraphics{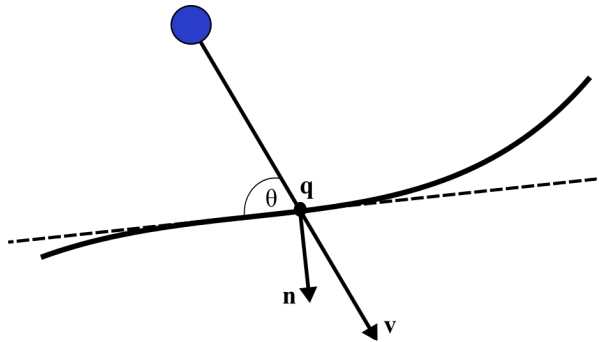} \caption{\label{billiard geo} 
Collision
geometry in a two-dimensional billiard. A particle with velocity $\mathbf{v}$
approaches the point $\mathbf{q}$ on the billiard boundary, where the outward
unit normal vector is $\mathbf{n}$. The dotted line represents the tangent line
to the boundary at $\mathbf{q}$} 
\end{figure} 
The biased distribution for each $\mathbf{q}$ can also be defined 
analogously to the one dimensional case, but we must also consider the collision
angle $\theta$, depicted for two-dimensional billiard in Fig.~\ref{billiard geo}. 
For a particle approaching the boundary
point $\mathbf{q}$ with speed $v$, $\theta$ is the angle between the particle's
velocity vector and the $d-1$ dimensional tangent surface to the wall at
$\mathbf{q}$, and $v \sin(\theta)$ thus gives the component of the particle's
velocity in the $\mathbf{\hat{n}}(\mathbf{q})$ direction. If the particle
collides when the wall has normal velocity $u$, then the relative speed of
approach just before the collision is determined by $v \sin(\theta)$ and $u$, so
$v \sin(\theta)$ determines the statistical bias towards collisions with large
negative $u$. We account for this by simply replacing $v$ with $v \sin(\theta)$
in Eq.~(\ref{P_v(u) 1d}), yielding 
\begin{equation} \label{P_v(u) nd}
P(u|v,\mathbf{q},\theta) = 
\begin{cases} P(u|0,\mathbf{q})\left(1-\frac{u}{v
\sin(\theta)}\right), & v \sin(\theta) \geq u_{max}(\mathbf{q}) \\ 
N(v,\theta)P(u|0, \mathbf{q})\left(1-\frac{u}{v
\sin(\theta)}\right)\Theta\boldsymbol{(}v \sin(\theta)-u\boldsymbol{)}, & v
\sin(\theta) < u_{max}(\mathbf{q}), 
\end{cases} 
\end{equation} 
Equation (\ref{P_v(u) nd}) determines the statistics of a particle's energy 
evolution in a quivering billiard. 
\\ 
\indent To summarize, we describe how one may construct
a quivering billiard and determine a particle's trajectory, without the need to
define a real, fully time-dependent billiard and take the quivering limit.
First, one must select a billiard shape by defining a surface $\{\mathbf{q} \}$,
then set boundary quivering by giving a value to $u_{c}$ and defining a scalar
field $h'(\mathbf{q},\Psi)$ on $\{\mathbf{q}\}$. If the constructed quivering
billiard is to honestly represent some deterministic billiard's motion in the
quivering limit, then $h'(\mathbf{q},\Psi)$ should be chosen to be a 
smooth function of $\mathbf{q}$ for any $\Psi$ wherever 
$\mathbf{\hat{n}}(\mathbf{q})$ in continuous. Using the field $h'$ and the 
value of $u_{c}$, one may then calculate the unbiased distribution 
$P(u|0,\mathbf{q})$ from Eq.~(\ref{P(u|0)nd}) for any $\mathbf{q}$ on the 
billiard boundary. For a particle in free flight inside the quivering 
billiard, the next collision location is found deterministically using the 
geometry of the billiard boundary, just as with a time-independent billiard. 
When a particle with velocity
$\mathbf{v}_{i}$ and speed $v_{i}$ collides with the boundary at $\mathbf{q}$
with a collision angle $\theta_{i}$, we draw a value of $u$ from the
distribution $P(u|v_{i},\mathbf{q}, \theta_{i})$. The particle's velocity
component tangent to the boundary remains constant, and the component normal to
the boundary just after the collision, $\mathbf{v}_{f} \cdot
\hat{\mathbf{n}}(\mathbf{q})$, is given by 
\begin{eqnarray}\label{v_f _nd}
\mathbf{v}_{f} \cdot \hat{\mathbf{n}}(\mathbf{q}) & = & 2u - \mathbf{v}_{i}
\cdot \hat{\mathbf{n}}(\mathbf{q}) \\ & = & 2u-v_{i}\sin(\theta_{i}). \nonumber
\end{eqnarray} 
The corresponding change in energy, $\Delta E$, is given by
\begin{equation} \label{Delta E nd_1} 
\Delta E = 2 m u^{2} - 2 m u\,v_{i}\sin\left(\theta_{i}\right). 
\end{equation} 
Analogously to the one dimensional case, if the selected value of $u$ is such 
that $0<u<v_{i} \sin(\theta_{i}) \leq 2u$, then a glancing collision occurs, 
and we draw a second value of $u$ using the same collision located and updated 
particle speed and collision angle, determined from Eqs.~(\ref{v_f _nd}) and 
(\ref{Delta E nd_1}).

\section{Energy Statistics} \label{sec:III} 
In this section, we study in detail
the statistical behavior of particles and ensembles in a $d$-dimensional
quivering billiard, with the aim of describing energy evolution of a ensemble of
initial conditions as a diffusion process. Our notation will be as follows:
$\mathbf{q}_{b}$ is the location of a particle's $b^{th}$ collision with the
billiard boundary, $\theta_{b}$ is the $b^{th}$ collision angle, $u_{b}$ is the
selected value of the wall velocity during the $b^{th}$ collision (sampled using
Eq.~(\ref{P_v(u) nd})), $v_{b-1}$ is the particle's speed just before the
$b^{th}$ collision, and $\Delta E_{b}$ is the change in particle energy due to
the $b^{th}$ collision, given by 
\begin{equation} \label{Delta E nd} 
\Delta E_{b} = 2 m u_{b}^{2} - 2 m u_{b}\,v_{b-1}\sin\left(\theta_{b}\right).
\end{equation} 
\indent In order to derive analytic results, we will assume that
the initial particle speeds $v_{0}$ are much larger than $u_{c}$, and we will
solve to leading order in the small parameter $\varepsilon = u_{c} / v_{0}$. We
regard $u_{c}$ as an $O(1)$ quantity, and $v_{0}$ as an $O(\varepsilon^{-1})$
quantity. This approximation allows us to ignore glancing collisions in our
analysis, and also allows us ignore the possibility of $v_{b-1} \sin(\theta_{b})
\leq u_{max}(\mathbf{q}_{b})$, so that the biased distributions at the time of
collision always take the form $P(u_{b}|v_{b-1}, \mathbf{q}_{b},\theta_{b}) =
P(u_{b}|0, \mathbf{q}_{b})\left(1-u_{b} / v_{b-1}\sin(\theta_{b})\right)$ (as
opposed to the more complicated Eq.~(\ref{P_v(u) nd})). The assumption
$\varepsilon \ll 1$ is not particularly restrictive; even if particles begin
with an initial speed comparable to or less than $u_{c}$, energy gaining
collisions are more likely than energy losing collisions due to the flux factor
in the biased distribution, and a slow particle will gain roughly $m u_{c}^{2}$
of energy during a collision according to Eq.~(\ref{Delta E nd}). Therefore, a 
slow particle will more than likely gain speed $u_{c} \sim O(1)$ during a 
single bounce, and after $1 / \delta$ bounces, where $\delta \ll 1$ is some 
small number, the particle will more than likely have a speed $v$ such that 
$u_{c} / v \lesssim \delta \ll 1$.  Thus, 
slow particles are very likely to eventually become fast particles, and the 
assumption 
$u_{c} / v \ll 1$ will give a better and better approximation over time. 
\\ 
\indent In the analysis, it will prove useful to consider both the
\textit{full dynamics} and \textit{frozen dynamics}, as is done in
Refs.~\cite{Jar1992, Jar1993}. If the frozen dynamics are used at the $b^{th}$
collision, the energy change $\Delta E_{b}$ is calculated, but the particle's
energy remains constant, and the angle of reflection is equal to the collision
angle $\theta_{b}$. In other words, the frozen dynamics are identical to those
of a time-independent billiard, but we calculate and keep track of the $\Delta
E_{b}$'s that would have occurred had the billiard walls been quivering. In the
full dynamics, the particle's energy is actually incremented by the calculated
value of $\Delta E_{b}$, and the angle of reflection is consequently altered.
\subsection{Expectations} \label{subsec:IIIA} 
Consider single a particle with
energy $E_{0}$ released at time $t_{0}$ in a $d$-dimensional quivering billiard.
The resulting particle trajectory generates a sequence of energy increments
$\{\Delta E_{1}, \Delta E_{2},...,\Delta E_{b-1}, \Delta E_{b}, \Delta
E_{b+1},... \}$. Let the operator $\{ ...\}_{b}$ denote the conditional
expectation value of the quantity $...$, given the outcomes of the previous
$b-1$ bounces. The first $b-1$ bounces determine $v_{b-1}$, $\mathbf{q}_{b}$,
and $\theta_{b}$, so the $b^{th}$ conditional expected energy change, $\mu_{b}
\equiv \{\Delta E_{b}\}_{b}$, can be calculated using the biased distribution
$P(u_{b}|v_{b-1}, \mathbf{q}_{b},\theta_{b})$ and the expression for $\Delta
E_{b}$ in Eq.~(\ref{Delta E nd}): 
\begin{eqnarray} \label{conditional_DE}
\mu_{b} & \equiv & \{\Delta E_{b} \}_{b} \\ 
& = & \int \mathrm{d}u_{b} P(u_{b}|v_{b-1},\mathbf{q}_{b},\theta_{b})
 \Delta E_{b} \nonumber \\ 
 & = & \int
\mathrm{d}u_{b} P(u_{b}|0,\mathbf{q}_{b}) \left(4 m u_{b}^{2} - \frac{2 m
u_{b}^{3}}{v_{b-1} \sin(\theta_{b})} - 2 m
u_{b}\,v_{b-1}\sin(\theta_{b})\right). \nonumber 
\end{eqnarray}
The integral in Eq.~(\ref{conditional_DE}) is taken over all possible values of 
$u_{b}$ at $\mathbf{q}_{b}$. 
\\ 
\indent Let $M_{n}(\mathbf{q}_{b})$ denote the 
$n^{th}$ moment of the wall velocity at $\mathbf{q}_{b}$ as measured by a 
stationary observer: 
\begin{equation} \label{moment_stationary} 
M_{n}(\mathbf{q}) = \int \mathrm{d}u P(u|0,\mathbf{q}) u^{n}. 
\end{equation} 
By construction,
$M_{1}(\mathbf{q}) = 0$ for all $\mathbf{q}$. Otherwise, $M_{n}(\mathbf{q}_{b})$
generally scales like $u_{c}^{n}$. The conditional mean thus simplifies to
\begin{eqnarray} \label{conditional_DE_1} 
\mu_{b} & = & \int \mathrm{d}u_{b}
P(u_{b}|0,\mathbf{q}_{b}) 4 m u_{b}^{2} \left( 1 - \frac{u_{b}}{2 v_{b-1}
\sin(\theta_{b})}\right) \\ 
& = & 4m M_{2}(\mathbf{q}_{b}) \left(1 -\frac{M_{3}(\mathbf{q}_{b}) /
M_{2}(\mathbf{q}_{b})}{2 v_{b-1}
\sin(\theta_{b})}\right). \nonumber 
\end{eqnarray} 
Similarly, the conditional variance $ \sigma_{b}^{2} $ is given by 
\begin{eqnarray} \label{conditional_var}
\sigma_{b}^{2} & \equiv & \{ (\Delta E_{b})^{2} \}_{b} - \{ \Delta E_{b}
\}^{2}_{b} \\ 
& = & \int \mathrm{d}u_{b}
P(u_{b}|v_{b-1},\mathbf{q}_{b},\theta_{b}) \left((\Delta E_{b})^{2} - \{ \Delta
E_{b} \}^{2}_{b}\right) \nonumber \\ 
& = & 4 m^{2} [M_{2}(\mathbf{q}_{b})]^{2}
\left( \frac{v_{b-1}^{2} \sin^{2}(\theta_{b})}{M_{2}(\mathbf{q}_{b})} -
3\frac{v_{b-1}\sin(\theta_{b})}{[M_{2}(\mathbf{q}_{b})]^{2} /
M_{3}(\mathbf{q}_{b})} +~3
\frac{M_{4}(\mathbf{q}_{b})}{[M_{2}(\mathbf{q}_{b})]^{2}} \right. \nonumber \\ 
& & ~\left. - 4 + \frac{ 4
M_{3}(\mathbf{q}_{b}) / M_{2}(\mathbf{q}_{b}) - M_{5}(\mathbf{q}_{b}) /
[M_{2}(\mathbf{q}_{b})]^{2}}{v_{b-1} \sin(\theta_{b})} -
\frac{[M_{3}(\mathbf{q}_{b})]^{2} / [M_{2}(\mathbf{q}_{b})]^{2}}{v_{b-1}^{2}
\sin^{2}(\theta_{b})} \right). \nonumber
\end{eqnarray} 
The terms enclosed in
the parentheses of Eqs.~(\ref{conditional_DE_1}) and (\ref{conditional_var}) are
ordered in increasing powers of $\varepsilon$. To leading order, we have
\begin{eqnarray} \label{conditional_leading} 
\mu_{b} = 4 m M_{2}(\mathbf{q}_{b}) \\ 
\sigma_{b}^{2} = 4 m^{2} M_{2}(\mathbf{q}_{b}) v_{b-1}^{2}
\sin^{2}(\theta_{b})\nonumber 
\end{eqnarray} 
The quantities $\mu_{b}$ and
$\sigma^{2}_{b}$ are $O(1)$ and $O\left( \varepsilon^{-2}\right)$, respectively;
average energy gain is moderate, and fluctuations are huge.

\subsection{Correlations} \label{subsec:IIIB} 
The conditional covariance between adjacent bounces, $\mathrm{Cov}_{b, b+1}$, 
is defined by 
\begin{eqnarray}\label{conditional_cov_def} 
\mathrm{Cov}_{b, b+1} & \equiv & \left\{\left(\Delta
E_{b}- \{\Delta E_{b} \}_{b}\right) \left(\Delta E_{b+1} - \{\Delta E_{b+1}
\}_{b}\right) \right\}_{b}\\
 & = & \{\Delta E_{b} \Delta E_{b+1} \}_{b} -
\{\Delta E_{b} \}_{b}\{\Delta E_{b+1} \}_{b}. \nonumber 
\end{eqnarray} 
The conditional expectations in Eq.~(\ref{conditional_cov_def}) are taken given 
the outcomes of the previous $b-1$ collisions, with the outcome of the $b^{th}$
collision yet to be determined. That is, we must average over all possible
realizations of the stochastic process $E_{b-1} \rightarrow E_{b-1} + \Delta
E_{b} \rightarrow E_{b-1} + \Delta E_{b} + \Delta E_{b+1}$, given the first
$b-1$ collisions. Denote $\{\Delta E_{b+1}| u_{b} \}_{b+1}$ as the conditional
expectation of $E_{b+1}$, given the first $b-1$ collision outcomes and supposing
that $u_{b}$ is the wall velocity during the $b^{th}$ collision. The expression
for $\{\Delta E_{b+1} \}_{b}$ is then 
\begin{equation} 
\{\Delta E_{b+1} \}_{b} =
\int \mathrm{d}u_{b} P(u_{b}|v_{b-1},\mathbf{q}_{b},\theta_{b}) \{\Delta
E_{b+1}| u_{b} \}_{b+1}. 
\end{equation} 
The expression for $\{\Delta E_{b} \Delta E_{b+1}\}_{b}$ can be written similarly: 
\begin{eqnarray} 
\{\Delta E_{b} \Delta E_{b+1} \}_{b} & = & \int
\mathrm{d}u_{b}\mathrm{d}u_{b+1}P(u_{b}|v_{b-1},\mathbf{q}_{b},\theta_{b})
P(u_{b+1}|v_{b},\mathbf{q}_{b+1},\theta_{b+1}|u_{b})\Delta
E_{b} \Delta E_{b+1} \\ 
& = & \int \mathrm{d}u_{b}
P(u_{b}|v_{b-1},\mathbf{q}_{b},\theta_{b}) \Delta E_{b} \{\Delta E_{b+1}| u_{b}
\}_{b+1}. \nonumber 
\end{eqnarray} 
The term $P(u_{b+1}|v_{b},\mathbf{q}_{b+1},\theta_{b+1}|u_{b})$ denotes the 
value of $P(u_{b+1}|v_{b},\mathbf{q}_{b+1},\theta_{b+1})$ 
when $v_{b}, \theta_{b+1}$, and $\mathbf{q}_{b+1}$ are determined given the 
first $b-1$ collision outcomes while supposing that $u_{b}$ is the wall 
velocity upon the $b^{th}$ collision. Equation~(\ref{conditional_cov_def}) 
can thus be expressed as 
\begin{equation}
\label{conditional_cov} \mathrm{Cov}_{b,b+1} = \int
\mathrm{d}u_{b}P(u_{b}|v_{b-1},\mathbf{q}_{b},\theta_{b}) \{\Delta E_{b+1}|
u_{b} \}_{b+1} \left(\Delta E_{b} - \{\Delta E_{b} \}_{b}\right).
\end{equation}
\indent If the frozen dynamics are used at the $b^{th}$ collision, then
$v_{b}$, $\theta_{b+1}$, and $\mathbf{q}_{b+1}$ are independent of $u_{b}$, so 
we have 
\begin{equation} \label{conditional _DE_Fr} \{\Delta E_{b+1}| u_{b}
\}_{b+1}|_{F} = \{\Delta E_{b+1} \}_{b+1}|_{F} = \mu_{b+1}|_{F}, 
\end{equation}
where $...|_{F}$ denotes the quantity $...$ evaluated using the frozen dynamics.
$\mu_{b+1}|_{F}$ carries no $u_{b}$ dependence, so it can be brought outside of
the integral in Eq.~(\ref{conditional_cov}), giving 
\begin{equation}
\mathrm{Cov}_{b,b+1}|_{F} = 0. 
\end{equation} 
Adjacent energy increments are thus statistically uncorrelated in the frozen 
dynamics. 
\\ 
\indent Under the assumption $\varepsilon \ll 1$, the frozen dynamics closely 
resemble the full dynamics over the time scale of a few bounces \cite{Jar1993}. 
Over such a time scale, we can regard the full dynamics trajectory as a 
stochastic perturbation of the deterministic frozen dynamics trajectory. Let 
$\mathbf{q}_{b+1}|u_{b} = \mathbf{q}_{b+1}|_{F} + \delta\mathbf{q}_{b+1}|u_{b}$ 
be the $(b+1)^{th}$ collision location when the full dynamics are used at the 
$b^{th}$ bounce, given the first $b-1$ collisions and supposing that $u_{b}$ is 
the observed wall velocity upon the $b^{th}$ collision. Equation 
(\ref{conditional_leading}) then gives, to leading order in $\varepsilon$ 
\begin{eqnarray} \label{DE_b+1}
\{\Delta E_{b+1}| u_{b} \}_{b+1} & = & 4 m M_{2}(\mathbf{q}_{b+1}|u_{b}) \\ & =
& 4 m M_{2}(\mathbf{q}_{b+1}|_{F}) + 4m \, \nabla M_{2}(\mathbf{q}_{b+1}|_{F})
\cdot \delta\mathbf{q}_{b+1}|u_{b}. \nonumber
 \end{eqnarray} 
where the gradient $\nabla M_{2}$ is constrained to act along directions 
tangent to the billiard boundary at $\mathbf{q}_{b+1}|_{F}$. In the appendix, 
we solve for $\|\delta\mathbf{q}_{b+1}|u_{b}\|$ to leading order in 
$\varepsilon$ and find
\begin{equation} \label{del_q_b+1} 
\|\delta\mathbf{q}_{b+1}|u_{b} \|= 2L_{b}|_{F} 
\frac{\cos(\theta_{b})}{\sin(\theta_{b+1}|_{F})} \frac{|u_{b}|}{v_{b-1}}, 
\end{equation} 
where $L_{b}|_{F}$  is the distance between the $b^{th}$ and $b+1^{th}$ 
collision locations in the frozen dynamics. Combining 
Eqs.~(\ref{conditional_cov}), (\ref{DE_b+1}), and (\ref{del_q_b+1}),
gives to leading order in $\varepsilon$ 
\begin{eqnarray}\label{conditional_cov_1st} 
\mathrm{Cov} _{b,b+1} & = & \int
\mathrm{d}u_{b}P(u_{b}|v_{b-1},\mathbf{q}_{b},\theta_{b}) \left(4m \, \nabla
M_{2}(\mathbf{q}_{b+1}|_{F}) \cdot \delta\mathbf{q}_{b+1}|u_{b} \right)
\left(\Delta E_{b} - \{\Delta E_{b} \}_{b}\right) \\ 
& = & \int \mathrm{d}u_{b}
P(u_{b}|0,\mathbf{q}_{b}) \left( 4 m\nabla M_{2}(\mathbf{q}_{b+1}|_{F}) \cdot
\frac{\delta\mathbf{q}_{b+1}|u_{b}}{\|\delta\mathbf{q}_{b+1}|u_{b} \|} \right.
\nonumber \\ & & \times~ \left. 2 L_{b}|_{F}
\frac{\cos(\theta_{b})}{\sin(\theta_{b+1}|_{F})}\frac{|u_{b}|}{v_{b-1}} \right)
\left(4 m u_{b}^{2} - 2m\frac{u_{b}^{3}}{v_{b-1}\sin(\theta_{b})} - 2 m u_{b}
v_{b-1} \sin(\theta_{b})\right. \nonumber \\ 
& & \left.-4 m
M_{2}(\mathbf{q}_{b}) + 4 m M_{2}(\mathbf{q}_{b}) \frac{u_{b}}{v_{b-1}
\sin(\theta_{b})} \right) \nonumber \\ 
& = & - 16
m^{2}L_{b}|_{F}\frac{\cos(\theta_{b})
\sin(\theta_{b})}{\sin(\theta_{b+1}|_{F})}\nabla M_{2}(\mathbf{q}_{b + 1}|_{F})
\cdot \int \mathrm{d}u_{b} P(u_{b}|0,\mathbf{q}_{b})
\frac{\delta\mathbf{q}_{b+1}|u_{b}}{\|\delta\mathbf{q}_{b+1}|u_{b} \|} u_{b}
|u_{b}| . \nonumber 
\end{eqnarray} 
All but the leading order terms are dropped
in the last line of Eq.~(\ref{conditional_cov_1st}). With exception to the
one-dimensional case, $\mathrm{Cov}_{b,b+1}$ is thus an $O(1)$ quantity. In a
one-dimensional billiard, the frozen and full dynamics always give the same
collision location, so $\{\Delta E_{b+1}| u_{b} \}_{b+1} = \{\Delta E_{b+1}
\}_{b+1}|_{F}$, and consequently, $\mathrm{Cov}_{b,b+1}$ is identically zero. \\
\indent The conditional correlation $\rho_{b,b+1}$ is defined as the normalized
conditional covariance, and is given by 
\begin{equation}
\label{conditional_corr} \rho_{b,b+1} = \frac{\mathrm{Cov}_{b,b+1}}{\sigma_{b}
\{\sigma_{b+1} \}_{b}}. 
\end{equation} 
To leading order in $\varepsilon$, the
conditional expectation $\{\sigma_{b+1} \}_{b}$ can be taken as the frozen
dynamics value in Eq.~(\ref{conditional_corr}). Therefore, the conditional
correlation $\rho_{b,b+1}$ is $O(\varepsilon^{2})$ (with exception to the
one-dimensional case, where $\rho_{b,b+1} = 0$). This quantity is very small,
and correlations between more distant collisions will further diminish due to
the mixing of particle trajectories induced by the stochastic wall motion. We
thus conclude that, in any dimension, correlations between energy increments
effectively decay over the time scale of a single collision.

\subsection{Ensemble averages} \label{subsec:IIIC} 
Consider now a microcanonical
ensemble of independent particles with energy $E_{0}$ released at time $t_{0}$.
The resulting trajectories will generate an ensemble of statistically
independent energy increment sequences, and we denote $\Delta E_{i, b}$ as the
$b^{th}$ recorded energy increment of the $i^{th}$ particle. Define the ensemble
averaged $b^{th}$ energy increment $\left< \Delta E_{b} \right>$ as
\begin{equation} \label{ensembleInc} 
\left< \Delta E_{b} \right> \underset{N
\rightarrow \infty}{=} \sum\limits_{i = 1}^{N} \frac{\Delta E_{i, b}}{N},
\end{equation} 
and the ensemble averaged $b^{th}$ conditional mean $\left<\mu_{b}\right>$ as 
\begin{equation} \label{ensembleExp}
\left<\mu_{b}\right> \underset{N \rightarrow \infty}{=} \sum\limits_{i = 1}^{N}
\frac{\mu_{i, b}}{N}, 
\end{equation}
where $\mu_{i,b} = \{\Delta E_{i, b} \}_{b}$. Equation 
(\ref{conditional_var}) shows that the $b^{th}$ conditional variances 
$\sigma_{i,b}^{2} \equiv \{\left(\Delta E_{i,b}\right)^{2} \}_{b} -
\{\Delta E_{i,b} \}^{2}$ are finite and bounded from above. Noting this, and the
fact that the series $\sum\limits_{k = 1}^{\infty} k^{-2}$ converges, we deduce
\begin{equation}\label{limit} 
\lim\limits_{N \rightarrow \infty}\sum\limits_{k =1}^{N} 
\frac{\sigma _{k, b}^{2}}{k^{2}} < \infty. 
\end{equation} 
By Kolmogorov's strong law of large numbers \cite{Revesz}, Eq.~(\ref{limit}) 
assures that, with probability unity, 
\begin{equation} \label{ensemb_ident1} \left<\Delta E_{b}
\right> = \left<\mu_{b} \right>. 
\end{equation} 
Combining Eqs.~(\ref{ensemb_ident1}),(\ref{ensembleExp}), 
and (\ref{conditional_leading}) gives, to leading order in $\varepsilon$, 
\begin{eqnarray} 
\left<\Delta E_{b} \right> & \underset{N \rightarrow \infty}{=} & 
\sum\limits_{i= 1}^{N} \frac{4m
M_{2}(\mathbf{q}_{i,b})}{N} \\ 
& = & 4m \left<M_{2}(\mathbf{q}_{b})\right>,
\nonumber 
\end{eqnarray} 
where $\mathbf{q}_{i,b}$ denotes the $b^{th}$ collision
location of the $i^{th}$ particle. By similar law of large number arguments, we
also have, to leading order in $\varepsilon$, 
\begin{equation}
\left<u_{b}^{2}\right> = \left<M_{2}(\mathbf{q}_{b})\right>, 
\end{equation}
where 
\begin{equation} \left<u_{b}^{2}\right> \underset{N \rightarrow \infty}{=}
\sum\limits_{i = 1}^{N} \frac{u_{i,b}^{2}}{N}, 
\end{equation}
and $u_{i,b}$ is the wall velocity during the $b^{th}$ collision of the 
$i^{th}$ particle. To leading order, we thus have 
\begin{equation} \left<\Delta E_{b}\right> =
4m\left<u_{b}^{2}\right>. 
\end{equation}

\subsection{Energy diffusion}\label{subsec:IIID} 
We now consider the normalized
energy distribution of an ensemble of independent particles, denoted by
$\eta(E,t)$. We have thus far shown that energy of any one ensemble member
evolves stochastically, in small increments, with correlations in energy changes
effectively decaying over a characteristic time scale given by time between
collisions. A particle's energy evolution is therefore effectively a Markov
process describing a random walk along an energy axis, so following
Refs.~\cite{Wilkinson1990, Jar1993}, we assert that $\eta(E,t)$ evolves like a
diffusion process and obeys a Fokker-Planck equation: 
\begin{equation}
\label{Fokker-Planck1} \partial_{t}\eta(E, t) =
-\partial_{E}\left[g_{1}(E,t)\eta(E, t)\right]
+\frac{1}{2}\partial^{2}_{E}\left[g_{2}(E,t)\eta(E, t)\right]. 
\end{equation}
The functions $g_{1}(E,t)$ and $g_{2}(E,t)$, the drift and diffusion terms,
respectively, are to be determined in this section. The energy of any one
particle in a quivering billiard evolves discretely in time, so the continuous
time evolution implied by Eq.~(\ref{Fokker-Planck1}) will be an accurate
description of the ensemble only down to a coarse-grained time scale. The time
scale must be large enough to ensure that most particles in the ensemble
experience at least a few bounces off the billiard wall, but small enough to
ensure the energy change experienced by most particles is small compared to
their total energy. Generally speaking, a diffusive description of a stochastic
process is only accurate over time scales larger than the process's typical
correlation time \cite{Jar1992, Jar1993}. We have established that energy
correlations for any one particle effectively decay over the time scale of a
single collision, thus, the diffusion approach to energy evolution in a
quivering billiard is justified on any time scale over which $\eta(E,t)$ can be
described by a continuous evolution. 
\\ 
\indent The drift term $g_{1}(E', t')$
is defined as the rate of ensemble averaged energy change for an ensemble of
particles all with energy $E'$ at time $t'$. Specifically, 
\begin{equation}\label{driftDef} 
g_{1}(E',t') \underset{\Delta t \rightarrow 0}{=}
\frac{\left<E(t' +\Delta t) - E(t') \right>}{\Delta t}, 
\end{equation} 
where $E(t') = E'$ for all particles in the ensemble, and $\left<E(t' + \Delta
t)\right>$ is the ensemble averaged particle energy at time $t' +\Delta t$. We
can not actually take the limit $\Delta t \rightarrow 0$ because $g_{1}$ has no
meaning over time scales for which the evolution of $\eta$ appears
discontinuous. Instead, we will let $\Delta t$ be the average time for which the
ensemble members make $B$ bounces after time $t'$, and we will find
corresponding ensemble averaged change in energy. We assume that $B$ is small
enough so that the particle energies change very little relative to $E'$ over
the time $\Delta t$, so that $\Delta t$ is the smallest coarse-grained time
scale for which Eq.~(\ref{Fokker-Planck1}) is valid for an ensemble with common
energy $E'$. We let $E_{B}$ be a particle's energy $B$ bounces after $t'$, and
find from Eq.~(\ref{ensemb_ident1}) 
\begin{eqnarray} \label{drift1} 
\left<E_{B} - E' \right> & = & \left< \sum\limits^{B} \Delta E_{b} \right> \\ 
& = & \sum\limits^{B} 4m \left<u_{b}^{2}\right>. \nonumber 
\end{eqnarray} 
We denote the coarse grained squared wall speed by $\overline{u^{2}}(t'; B)$, 
defined as the time average of $\left< u^{2}_{b} \right>$ over the first $B$ 
bounces after $t'$: 
\begin{equation} \label{coarseVel} 
\overline{u^{2}}(t'; B) = \sum\limits^{B} \frac{\left<u^{2}_{b}\right>}{B}. 
\end{equation} 
We thus have
\begin{equation} \label{DeltaEB} 
\left<E_{B} - E' \right> = 4 m~ \overline{u^{2}}(t';B) B 
\end{equation} 
\indent The time scale $\Delta t$ corresponding to the $B$ bounces after 
$t'$ is the ensemble averaged total free flight time over which the $B$ 
bounces occur. If we denote by $\Delta t_{b}$ a particle's $b^{th}$ free 
flight time after $t'$, we have 
\begin{equation}\label{deltaT} 
\Delta t =\sum\limits^{B} \left<\Delta t_{b}\right>.
\end{equation} 
We are assuming small wall velocities, so the particles' speeds
change very little relative to their initial speed $\sqrt{2E'/m}$ over the $B$
bounces. Therefore, to leading order in $\varepsilon$, we have 
\begin{equation}
\label{deltaT2} \Delta t_{b} = \sqrt{\frac{m}{2E'}} l_{b}, 
\end{equation} 
where $l_{b}$ denotes a particles $b^{th}$ free flight distance after $t'$. We 
now define the coarse grained free flight distance, $\overline{l}(t',B)$ by time
averaging the ensemble average of $l_{b}$ over the first $B$ bounces after $t'$:
\begin{equation} \label{l_B} 
\overline{l}(t';B) = \sum\limits^{B}\frac{\left<l_{b}\right>}{B} 
\end{equation}
Substituting Eqs.~(\ref{l_B}) and (\ref{deltaT2}) into Eq.~(\ref{deltaT}) gives
\begin{equation}\label{deltaT3} \Delta t =
B~\overline{l}(t';B)\sqrt{\frac{m}{2E'}}, 
\end{equation} 
and substituting for $B$ in Eq.~(\ref{DeltaEB}) gives 
\begin{equation} \label{DeltaEB2}
\left<E_{B} - E' \right> = \Delta t~ \frac{4 \sqrt{2m}~ \overline{u^{2}}(t';B)}
{\overline{l}(t';B)} E'^{\frac{1}{2}}. 
\end{equation}
Equation (\ref{DeltaEB2}) gives the ensemble averaged change in energy over the
time $\Delta t$ after $t'$ for an ensemble of particles with energy $E'$.
Comparing to Eq.~(\ref{driftDef}), we see that dividing both sides of
Eq.~(\ref{DeltaEB2}) by $\Delta t$ gives us $g_{1}(E',t')$. We thus have,
\begin{equation} \label{g1_func} 
g_{1}(E, t) = \frac{4 \sqrt{2m}~ \overline{u^{2}}(t)}{\overline{l}(t)} 
E^{\frac{1}{2}}, 
\end{equation} 
where we have switched from primed to unprimed variables, and the dependence 
on $B$ has been suppressed. 
\\ 
\indent The diffusion term $g_{2}(E',t')$ is defined as
\begin{equation} \label{driffDef} g_{2}(E',t') \underset{\Delta t \rightarrow
0}{=} \frac{\left<\left(E(t' +\Delta t) - E(t')\right)^{2} \right>}{\Delta t},
\end{equation} 
where $E(t') = E'$ for all particles in the ensemble, and 
$\left<E(t' + \Delta t)\right>$  is the ensemble averaged particle energy at time
$t' +\Delta t$. An expression for the diffusion term can be found by employing
similar methods used to find the drift term. Alternatively, $g_{2}(E,t)$ can be
found by invoking Liouville's theorem, as in Ref.~\cite{Jar1992}. Combing
Liouville's theorem and the Fokker-Planck equation allows one to deduce a
fluctuation-dissipation relation: 
\begin{equation} \label{FlucDiss1} 
g_{1}(E, t) = \frac{1}{2 \Sigma(E)} \partial_{E}\left[\Sigma(E)g_{2}(E, t)
\right],
\end{equation} 
where $\Sigma(E)$ is the microcanonical partition function of a
single particle with energy $E$ in the corresponding frozen billiard. In a $d$
dimensional billiard, the microcanonical partition function is given by
\cite{Jar1993} 
\begin{equation} \label{SigmaDef} 
\Sigma(E) = \frac{1}{2}V_{d}\,\Omega_{d}\left(2m\right)^{\frac{d}{2}}
 E^{\frac{d}{2} - 1},
\end{equation} 
where $\Omega_{d}$ is the $d$-dimensional solid angle, and
$V_{d}$ is the $d$-dimensional billiard's volume. Combining
Eqs.~(\ref{g1_func}),(\ref{FlucDiss1}), and (\ref{SigmaDef}), we find
\begin{equation} \label{g2_func} g_{2}(E, t) = \frac{4}{d+1}\frac{4 \sqrt{2m}~
\overline{u^{2}}(t)}{\overline{l}(t)} E^{\frac{3}{2}}. 
\end{equation} 
This method of determining $g_{2}$ allows for an additive constant, but this 
constant must be identically zero; when $E = 0$, the particles are motionless and 
there can be no drift or diffusion of energies, so we must have 
$g_{1}(0, t)= g_{2}(0, t) = 0$. 
\\ 
\indent With our expressions for $g_{1}$ and $g_{2}$, we may rewrite
the Fokker-Planck equation: 
\begin{equation} \label{Fokker-Planck2}
\partial_{t}\eta(E, t) = \frac{2\alpha(t)}{d+1} \partial_{E}\left[
E^{\frac{1+d}{2}}\partial_{E}\left(E^{\frac{2-d}{2}} \eta(E, t)\right)\right]
\end{equation} 
where we define $\alpha(t)$ as 
\begin{equation} \label{alpha def}
\alpha(t) \equiv \frac{4 \sqrt{2m}~ \overline{u^{2}}(t)}{\overline{l}(t)}.
\end{equation} 
Equation~(\ref{Fokker-Planck2}) can be simplified by defining a
rescaled time $s$: 
\begin{equation} \label{s} 
s = \int_{t_{0}}^{t} \mathrm{d}t'\alpha(t'), 
\end{equation} 
which gives 
\begin{equation} \label{Fokker-Planck3}
\partial_{s}\eta(E, s) = \frac{2}{d+1} \partial_{E}\left[
E^{\frac{1+d}{2}}\partial_{E}\left(E^{\frac{2-d}{2}} \eta(E, s)\right)\right]
\end{equation} 
\indent Equation~(\ref{Fokker-Planck3}) can be solved by
separation of variables. We assume a solution of the form $\phi(s) f(E)$, and
upon making the substitutions $F(E) = E^{\frac{3- d}{4}} f(E)$ and $z =
E^{\frac{1}{4}}$ one finds a first order homogeneous linear differential
equation for $\phi(s)$ and a Bessel equation of order $d-1$ for $F(z)$. The
details of the separation of variables, including existence, uniqueness, and
boundary conditions, are given in Ref.~\cite{KACK2013} and will be omitted 
here.
We also acknowledge a similar, much older, one-dimensional solution given in
Ref.~\cite{Z&C}. The separation of variables solution is 
\begin{equation}
\label{FPsoln1} \eta(E, s) = E^{\frac{d-3}{4}} \int_{0}^{\infty} \mathrm{d}k \,
A(k) J_{d-1}(k E^{\frac{1}{4}}) e^{-\frac{s k^{2}}{8(d + 1)}}, 
\end{equation}
where $J_{d-1}$ is an ordinary Bessel function of order $d-1$, and the
amplitudes $A(k)$ are found by taking a Hankel transform of the initial ensemble
$\eta(E, 0)$. When the ensemble begins in the microcanonical distribution with
energy $E_{0}$, we have $\eta(E, 0) = \delta(E- E_{0})$, and a closed form
expression for $A(k)$ results. The energy distribution $\eta(E, s)$, subject to
$\eta(E, 0) = \delta(E- E_{0})$, is then 
\begin{equation} \label{FPsoln2}
\eta(E, s) =
\frac{1}{4E_{0}^{\frac{1}{2}}}\left(\frac{E}{E_{0}}\right)^{\frac{d-3}{4}}
\int_0^{\infty} \mathrm{d}k \, k J_{d-1}(k E_{0}^{\frac{1}{4}})J_{d-1}(k
E^{\frac{1}{4}}) e^{-\frac{s k^{2}}{8(d + 1)}}. 
\end{equation} 
Making use of an identity of Bessel integrals utilized in Eq.~(22) of 
Ref.~\cite{KACK2013}, we can solve the integral in Eq.~(\ref{FPsoln2}) 
and simplify the expression to
\begin{equation}\label{FPsoln3} 
\eta(E, s) =
\frac{d+1}{sE_{0}^{\frac{1}{2}}}\left(\frac{E}{E_{0}}\right)^{\frac{d-3}{4}}
I_{d-1}\left[\frac{4(d+1)}{s}E_{0}^{\frac{1}{4}} E^{\frac{1}{4}}\right]
e^{-\frac{2(d+1)}{s}\left(E_{0}^{\frac{1}{2}} + E^{\frac{1}{2}}\right)},
\end{equation} 
where $I_{d-1}$ is a modified Bessel function of order $d-1$. Using this energy
distribution, we can find the ensemble averaged energy as a function
of time: 
\begin{equation}\label{<E>} 
\left<E(s)\right> = \frac{d}{d+1}\frac{s^{2}}{4}+\sqrt{E_{0}}~s + E_{0}. 
\end{equation}  
\indent
Equation (\ref{FPsoln3}) is only valid under the assumption $\varepsilon \ll 1$.
If we begin with an ensemble where $\varepsilon$ is order unity or larger, over
sufficiently long time, the slow particles inevitably gain so much energy that
the fast particle assumption holds and Eq.~(\ref{FPsoln3}) becomes valid
asymptotically. We can thus find a universally valid asymptotic energy
distribution by considering Eq.~(\ref{FPsoln2}) or Eq.~(\ref{FPsoln3}) in the
limit of very large $s$. Specifically, if $k \ll d / \sqrt{E_{0}}$ for all
$k^{2} \gg 8(d+1) / s$, which implies that $s \gg 8 \sqrt{E_{0}}(d+1) / d$, one
can approximate $J_{d-1}(k E_{0}^{\frac{1}{4}})$ by the lowest order term in its
Taylor expansion over the non-negligible contributions to the integral in
Eq.~(\ref{FPsoln2}), and the solution reduces to 
\begin{equation} \label{AssSoln2} 
\eta_{a}(E, s) = \frac{1}{2E\Gamma(d)}\left[\frac{2(d+1)}{s}
E^{\frac{1}{2}}\right]^{d} e^{-\frac{2(d+1)}{s}E^{\frac{1}{2}}}, 
\end{equation}
where $\Gamma$ is the gamma function. One can easily verify that $\eta_{a}(E,s)$
is normalized and obeys the Fokker-Planck equation. Using the asymptotic energy
distribution Eq.~(\ref{AssSoln2}), we find the ensemble averaged energy at a
large times to be 
\begin{equation} \label{<E>Ass} 
\left<E(s)\right>_{a} = \frac{d}{1+d} \frac{s^{2}}{4}. 
\end{equation}
\indent The results of this
section are summarized as follows. In the quivering limit, correlations in
particle energy decay over the time scale of a single collision, and as a
result, the energy distribution of an ensemble evolves diffusively, regardless
of the shape and dimensionality of the billiard boundary. Ensembles universally
evolve to the asymptotic energy distribution given in Eq.~(\ref{AssSoln2}), and
ensemble averaged energy asymptotically grows quadratically in time. Before
discussing the implications and broader context of these results, we comment on
the interpretations of the coarse grained quantities $\overline{l}$ and
$\overline{u^{2}}$. 
\\ 
\indent If the particular billiard shape is ergodic, then
their exists a characteristic ergodic time scale over which ensembles uniformly
explore the entire billiard boundary. Invoking ergodicity and replacing time
averages with phase space averages, we deduce that, over time scales greater
than the ergodic time scale, $\overline{l}$ will be the billiard's mean free
path, and $\overline{u^{2}}$ will be the second wall moment $M_{2}(\mathbf{q})$
uniformly averaged over the billiard boundary. This implies that, over time
scales greater than the ergodic time scale, $g_{1}$ and $g_{2}$ are
time-independent and that $\alpha$ is merely a constant. In this case, the
expression for $g_{1}$ in Eq.~(\ref{g1_func}) is equivalent to the wall 
formula,
which was originally used to model energy dissipation from collective to
microscopic degrees of freedom in nuclear dynamics \cite{JarSwia1993}. In
non-ergodic billiards, or over time scales shorter than the ergodic time scale
in ergodic billiards, $\overline{l}$ and $\overline{u^{2}}$ will generally be
time-dependent and can not be interpreted in terms of properties of the 
billiard
shape alone. Nevertheless, they are still well-defined properties of the
ensemble; $\overline{l}$ is simply the ensemble's average free flight
distance over the coarse grained time scale, and $\overline{u^{2}}$ is the
average squared wall velocity for the collisions taking place over the coarse
grained time scale.

\section{Discussion} \label{sec:IV}

\subsection{Approximate Quivering}\label{ApproxQuiv} 
The quivering limit is most
certainly an idealization of time-dependent billiard motion; no real billiard
boundary can actually move with zero amplitude and period. However, if the
idealized system is defined in a physically consistent manner, then we expect
that for smaller and smaller $a$ and $\tau$, real time-dependent billiards will
be better and better approximated by quivering billiards. We now clarify how
small $a$ and $\tau$ must actually be for a time-dependent billiard to be
well-approximated by a quivering billiard. 
\\ 
\indent In Refs.~\cite{L&L1972}
and \cite{LLC1980}, Lieberman, Lichtenberg, and Cohen studied the Fermi-Ulam
model numerically and analytically using dynamical systems theory. It was shown
that the energy evolution of a particle in the Fermi-Ulam model is generically
diffusive and can be described by a Fokker-Planck equation for particle speeds
such that, using our notation from Sec.~\ref{sec:IIA}, $v \ll u_{c} \sqrt{L /
a}$. The value $u_{c} \sqrt{L / a}$ is associated with the stability of periodic
orbits in $v$-$\Psi$ space, where $v$ and $\Psi$ are the particle velocity and
wall phase during collisions, respectively. At particle speeds much below $u_{c}
\sqrt{L / a}$, Refs.~\cite{L&L1972} and \cite{LLC1980} show that periodic orbits
in $v$-$\Psi$ space are unstable, dynamical correlations are small, and
trajectories in $v$-$\Psi$ space are generally chaotic (the language of the day
labelled such trajectories stochastic as opposed to chaotic). At particle speeds
above $u_{c} \sqrt{L / a}$, periodic orbits begin to stabilize, correlations
become important, and the presence of elliptic islands and invariant spanning
curves inhibit energy growth \cite{L&L1972} \cite{LLC1980}. In a one-dimensional
quivering billiard, correlations vanish, trajectories are stochastic, and
particle energy evolves diffusively, so, based on Lieberman, Lichtenberg, and
Cohen's work, we see that a quivering billiard is a good description of the
Fermi-Ulam model when $v \ll u_{c} \sqrt{L / a}$. As $a$ becomes smaller and
smaller with $u_{c}$ held fixed, elliptic islands and invariant spanning curves
move away to regions of larger and larger particle speeds, correlations become
smaller and smaller due to the more and more erratic wall motion, and quivering
becomes a valid approximation for wider and wider ranges of particle speeds. As
$a$ approaches zero in the idealized limit, the infinitely erratic wall motion
destroys correlations, elliptic islands and spanning curves occur only at
infinite energy, and quivering becomes an exact description for all particle
speeds. The same reasoning can be applied to higher dimensional time-dependent
billiards; as $a$ becomes smaller and smaller with $u_{c}$ held constant,
correlations become smaller and smaller and non-diffusive dynamics occur at
higher and higher energies. We thus claim that when $v \ll u_{c}\sqrt{l_{c} /
a}$ for all possible particle speeds $v$ that could be observed in a simulation
or experiment, where $l_{c}$ is a characteristic free-flight distance, an
arbitrary-dimensional time-dependent billiard will be approximately a quivering
billiard. 
\\ 
\indent Due to the inevitable increase in particle energy, the
speed bound inequality $v \ll u_{c}\sqrt{l_{c} / a}$ implies that quivering 
will closely approximate a real billiard simulation or experiment 
only up to some maximum time $t_{max}$. The value of
$t_{max}$ depends on the particles' initial energy distribution, but we can 
estimate its scaling behavior in situations where the actual energy 
distribution is able to evolve the asymptotic distribution given in 
Eq.~(\ref{AssSoln2}). In such cases, the average particle speed at large 
times can be estimated from the asymptotic 
ensemble averaged energy given by Eq.~(\ref{<E>Ass}), and we find 
$v \sim t ~ u_{c}^{2} / l_{c}$. Substituting this estimate for $v$ into the 
speed bound inequality yields 
$t \ll(l_{c} / a)^{1/2} ~ (l_{c} /u_{c}) = (l_{c}/a)^{3/2} \tau$. We thus have
$t_{max} \sim (l_{c} / a)^{1/2} ~ (l_{c} /u_{c}) = (l_{c}/a)^{3/2} \tau$. As
expected, in the quivering limit, $t_{max}$ diverges.

\subsection{Consistency}\label{consist} 
Quivering wall motion corresponds to
volume preserving billiard motion with negligible correlations in particles'
energy changes. Therefore, if the quivering limit is actually physically
meaningful, then the results obtained in Sec.~\ref{sec:III} should agree with
previous time-dependent billiard literature for the special case of volume
preserving billiard motion with negligible correlations in energy changes. We
now highlight three such examples. 
\\ 
\indent In Ref.~\cite{LLC1980},
$\left<\Delta E\right>$ and $\left<(\Delta E)^{2}\right>$ are calculated for a
single collision in the Fermi-Ulam model, assuming periodic wall motion (which
corresponds to volume preserving billiard motion on average) and no correlations
in the wall velocity between collisions. The authors also assume, without
explicitly stating, that the wall velocity is an even function of time. The
expressions obtained in Ref.~\cite{LLC1980} are in fact identical to our
expressions for $\{\Delta E_{b}\}_{b}$ in Eq.~(\ref{conditional_DE_1}) and
$\{(\Delta E_{b})^{2}\}_{b}$, which can be found by adding $\{\Delta
E_{b}\}_{b}^{2}$ to Eq.~(\ref{conditional_var}), under the assumption that all
odd moments of the wall velocity $M_{2n + 1}$ vanish. The odd moments vanish in
a quivering billiard when we take the quivering limit of wall motion defined by
an even function of time, so our results agree perfectly with those of
Ref.~\cite{LLC1980}. 
\\ 
\indent Reference \cite{Jar1993} studies the energy
evolution of ensembles of independent particles in chaotic adiabatic billiards
in two and three dimensions. A Fokker-Planck equation to describe the evolution
of the energy distribution is proposed, and expressions for the corresponding
drift and diffusion coefficients are derived. These results are obtained for
general adiabatic billiard motion, under the assumption that correlations in a
particle's energy changes decay over the mixing time scales corresponding to the
frozen chaotic billiard shapes. The expressions for $g_{1}$ and $g_{2}$ are
given in terms of a diffusion constant $D$, and an explicit expression for $D$
is given using the \textit{quasilinear approximation} - the assumption that
energy changes between bounces are completely uncorrelated. Under the
quasilinear approximation, assuming volume preserving billiard motion, the
expressions for $g_{1}$ and $g_{2}$ in Ref.~\cite{Jar1993} are identical to our
two and three-dimensional expressions for $g_{1}$ and $g_{2}$ in
Eqs.~(\ref{g1_func}) and (\ref{g2_func}), respectively, for ergodic billiards,
over time scales greater than the ergodic time scale. Our results are thus
consistent with those of Ref.~\cite{Jar1993}. It is remarked in
Ref.~\cite{Jar1993} that it is not precisely clear under what conditions the
quasilinear approximation will be valid for time-dependent billiards in 
general, but roughly speaking, the approximation requires the billiard shapes 
and motion to be ``sufficiently irregular.'' Our results help clarify this 
issue; the quasilinear approximation is justified when a time-dependent 
billiard is approximately quivering, and the quasilinear approximation is in 
fact exact, not an approximation, in the quivering limit. 
\\ 
\indent In Ref.~\cite{JarSwia1993},
it is shown that the velocity distribution for independent particles in a
time-dependent irregular container is asymptotically universally an 
exponential. This work assumes an isotropic velocity distribution, volume 
preserving billiard motion, and a three-dimensional billiard. If we assume an 
isotropic velocity distribution in a quivering billiard, we can change 
variables from energy to velocity in Eq.~(\ref{AssSoln2}), and we find the 
asymptotic velocity distribution $f_{a}(\mathbf{v}, s)$ in arbitrary dimensions
\begin{equation}\label{assVelSol} 
f_{a}(\mathbf{v}, s) =
\frac{1}{\Omega_{d}\Gamma(d)}\left(\frac{2(d+1)}{s}\sqrt{\frac{m}{2}}\right)^
{d}e^{-\frac{2(d+1)}{s}||\mathbf{v}||}.
\end{equation} 
In agreement with Ref.~\cite{JarSwia1993}, the isotropic velocity
distribution in a quivering billiard is universally an exponential in all
dimensions. For a three-dimensional chaotic quivering billiard, where $s=\alpha
t$ and chaotic mixing ensures an isotropic velocity distribution,
Eq.~(\ref{assVelSol}) is identical to the velocity distribution obtained in
Ref.~\cite{JarSwia1993}.

\subsection{Fermi acceleration} \label{subsec:IVB} 
Equation (\ref{<E>Ass}) shows
that the ensemble averaged growth is unbounded, increasing quadratically in
time. Unbounded average energy growth in time-dependent billiards is known as
Fermi acceleration. Fermi acceleration was originally proposed by Fermi as the
mechanism by which cosmic rays gain enormous energies through reflections off 
of moving magnetic fields \cite{Fermi1949}, and since become an active field of
research in its own right. The current research generally seeks to determine
under what conditions time-dependent billiards allow for Fermi acceleration, 
and to understand how the dynamics of sequence of frozen billiard shapes 
affects the energy growth rate. In Refs.~\cite{Z&C, Brahic, L&L1972, LLC1980}, 
it was established that sufficiently smooth wall motion in the one-dimensional
Fermi-Ulam model prohibits Fermi acceleration, and that non-smooth wall motion 
allows for Fermi acceleration that may be much slower than quadratic in time. 
While the one-dimensional billiard is always integrable, higher dimensional 
billiards allow for integrable, pseudo-integrable, chaotic, or mixed dynamics. 
In Ref.~\cite{LRA2000}, it was conjectured that fully chaotic frozen billiard
shapes are a sufficient condition for Fermi acceleration in multi-dimensional
time-dependent billiards, and the energy growth rate in such billiards was
thought to be quadratic in time \cite{LRA2000, GRT2012}. It has since been 
shown that the problem is a bit more subtle; certain symmetries in the 
sequence frozen billiard shapes can prohibit or stunt the quadratic energy 
growth in chaotic billiards \cite{BAT2014A}. The problem is complicated for 
non-chaotic multi-dimensional billiards as well. Integrable billiards may 
prohibit \cite{KdC1999} or allow \cite{LPKDS2011} quadratic or slower Fermi 
acceleration, while exponential Fermi acceleration is possible for
pseudo-integrable billiards \cite{Shah2011} and billiards with multiple 
ergodic components \cite{STRK2010, GRKST2011, GRT2012, GRKT2014, BAT2014B} 
with possibly mixed or pseudo-integrable dynamics.
\\ 
\indent Given the complexities observed in the previous literature, our result 
in Eq.~(\ref{<E>Ass}) is surprising; in the quivering limit, regardless of the
dimensionality or underlying frozen dynamics, time-dependent billiards
universally show quadratic Fermi acceleration. The apparent contradiction
between our work and previous work is due to a difference in the limits 
studied. Both our work and the previous literature, because of the inevitable 
speed up of particles, analyze time-dependent billiards in the adiabatic 
limit, where the wall speed is much slower than the particle speed. In the 
previous literature, however, the period of billiard oscillations is typically 
fixed and non-zero (with numerical results often presented as a function of 
the oscillation amplitude), so in the adiabatic limit, the typical time 
between collisions is always much shorter than the billiard's oscillation 
period. In our work, the oscillation period approaches zero, so the time 
between collisions is always much larger than the oscillation period, even in 
the adiabatic limit where particles move much faster than walls.

\subsection{Fixed wall simplifications} \label{subsec:IVC} 

\indent An
alternative simplification similar to the quivering billiard has been 
frequently employed in the literature. The so-called static wall approximation 
(sometimes called the simplified Fermi-Ulam model) was originally introduced in
\cite{L&L1972} in order to ease the analytical and numerical study of the
Fermi-Ulam model, and through the years has become a standard approximation
assumed valid for small oscillation amplitudes, often studied entirely in lieu
of the exact dynamics. See Ref.~\cite{L&L1972, LLC1980, LRA1999, LRA2000,
LMS2004, KPDC2006, KPDC2007, KDC2008} for example. Using the notation of
Sec.~\ref{sec:II}, assuming $v \gg u_{c}$ so that we may ignore glancing
collisions for the sake of simplicity, the dynamics of the one-dimensional
Fermi-Ulam model can be described by the deterministic map, 
\begin{subequations} \label{FUmap} 
\begin{eqnarray} 
v_{b} & = & v_{b-1} - 2 u(t_b), \label{FUmapv} \\
t_{b} & = & t_{b-1} + \frac{2 L}{v_{b-1}} +\frac{g(t_{b}) +
g(t_{b-1})}{v_{b-1}}, \label{FUmapt} 
\end{eqnarray}
\end{subequations} 
while the corresponding static wall approximation is given by the 
deterministic map,
\begin{subequations} \label{SWmap} 
\begin{eqnarray} 
v_{b} & = & v_{b-1} - 2 u(t_b), \label{SWmapv} \\ 
t_{b} & = & t_{b-1} + \frac{2 L}{v_{b-1}}. \label{SWmapt} 
\end{eqnarray} 
\end{subequations} 
In the above maps, $v_{b-1}$ is
the particle's velocity just before the $b^{th}$collision, and $t_{b}$ is the
time of the $b^{th}$ collision. An analogous static wall approximation can be
constructed for higher dimensional billiards \cite{LRA1999, LRA2000, KPDC2007}.
Like the quivering billiard, the static wall approximation eliminates the
implicit equations for the time between collisions by holding the billiard
boundary fixed. The two models differ because the static wall approximation
assumes $u(t_{b})$ to be a well behaved function. It is common practice to
consider stochastic versions of the maps (\ref{FUmap}) and (\ref{SWmap}), where
$u(t_{b})$ is replaced by $u(t_{b}+\zeta)$ for some random variable $\zeta$
~\cite{L&L1972, LRA1999, LRA2000, KPDC2006, KPDC2007,KDC2008}. The stochastic
case simulates the effects of external noise on the system and allows one to
average over $\zeta$ when determining ensemble averages, which often 
facilitates analytical calculations. 
\\ 
\indent In Refs.~\cite{KPDC2006, KPDC2007}, Karlis
et al. show that the stochastic static wall map and its analogue for the
two-dimensional Lorentz gas give one half the asymptotic energy growth rate of
the stochastic Fermi-Ulam map. This inconsistency exists even for small $a$, so
Karlis et al. conclude that (\ref{SWmap}) is not a valid approximation of
(\ref{FUmap}). We add that the same factor of two discrepancy can be observed
between our quivering billiard expression for $g_{1}$ and the corresponding
expressions obtained from the deterministic static wall maps given in
\cite{L&L1972, LLC1980, LRA1999, LRA2000}. In an early study of the Fermi-Ulam
model, Ref.~\cite{Z&C} obtains a drift term that is actually in agreement with
the static wall approximation value, but a careful reading reveals that the
authors make a series of simplifications that inadvertently reduce their
Fermi-Ulam model to the static wall approximation. Ref.~\cite{KPDC2006} 
corrects for the energy inconsistency to a high degree of accuracy in the 
stochastic case by introducing the hopping wall approximation. The hopping 
wall approximation assumes wall motion slow enough such that the moving wall's 
position at the
$b^{th}$ bounce can be approximated by its position at the $(b-1)^{th}$ bounce,
or by its position at the time of the particle's collision with the fixed wall
just after the $(b-1)^{th}$ bounce. This approximation allows $g(t_{b})$ in
Eq.~(\ref{FUmapt}) to replaced by either $g(t_{b-1})$ or $g(t_{b-1} + L /
v_{b-1})$. An analogous hopping wall approximation for two dimensions is
presented in \cite{KPDC2007}. Like the static wall approximation, the hopping
wall approximation eliminates the implicit equations for the time between
collisions, which eases numerical and analytical study. Based on the hopping
wall approximation's more accurate asymptotic energy growth rate, Karlis et al.
conclude in Refs.~\cite{KPDC2006, KPDC2007} that the energy discrepancy between
the Fermi-Ulam model and the static wall approximation is due to dynamical
correlations induced by small changes in the free flight time between 
collisions which are neglected in the static wall approximation. 
\\ 
\indent Based on the
results of this paper, we propose an alternative explanation of the energy
discrepancy. The energy discrepancy is observed because the static wall
approximation is simply unphysical, and it can not accommodate for the fact
that, due to the relative motion between the particles and walls, collisions
with inward moving walls are more likely than collisions with outward moving
walls. In fact, defining the quivering billiard \textit{without} the flux factor
in the biased distribution (so that the biased and unbiased distribution are
equal) reproduces the asymptotic energy growth rate predicted by the stochastic
static wall approximation. Evidently, the last term in Eq.~(\ref{FUmapt}) is
responsible for the bias towards inward moving wall collisions in the exact
Fermi-Ulam model, and hopping wall approximation's estimate of this term is
responsible for its more accurate energy growth rate. Although the static wall
approximation is a mathematically well-defined dynamical system, it is an
ill-posed physical system for the following reasons. If a billiard boundary is
truly static such that (\ref{FUmapt}) somehow reduces to (\ref{SWmapt}), then 
we must have $a \rightarrow 0$. But if $a \rightarrow 0$, then $u_{c} 
\rightarrow 0$ and the billiard becomes trivially time-independent unless 
$\tau \rightarrow 0$ as well. However, if both $a$ and $\tau \rightarrow 0$, 
then $u(t)$ can not
be a well-behaved function as required by the definition of the static wall 
map, and, as argued in Sec.~\ref{sec:II}, the wall velocity becomes 
stochastic. This logic seems to be unavoidable; if the walls are to be 
genuinely fixed, then physical consistency demands that the wall motion must 
be non-existent or stochastic. Based on this reasoning, we propose the 
following conjecture: any physically consistent, non-trivial, fixed wall 
limit of a time-dependent billiard must be physically equivalent to the 
quivering limit, and the corresponding quivering billiard as defined in this 
paper yields the correct dynamics and energy growth rate (by physically 
equivalent, we mean equivalent energy and velocity statistics). Of 
particular note, corrections to the free flight time between collisions are 
not needed to achieve the correct energy growth rate.

\section{Examples and Numerics} \label{sec:V} 
\begin{figure}[b]
\subfloat[$t=100$] {\label{FUM1D a} \includegraphics[width=4.0cm] 
{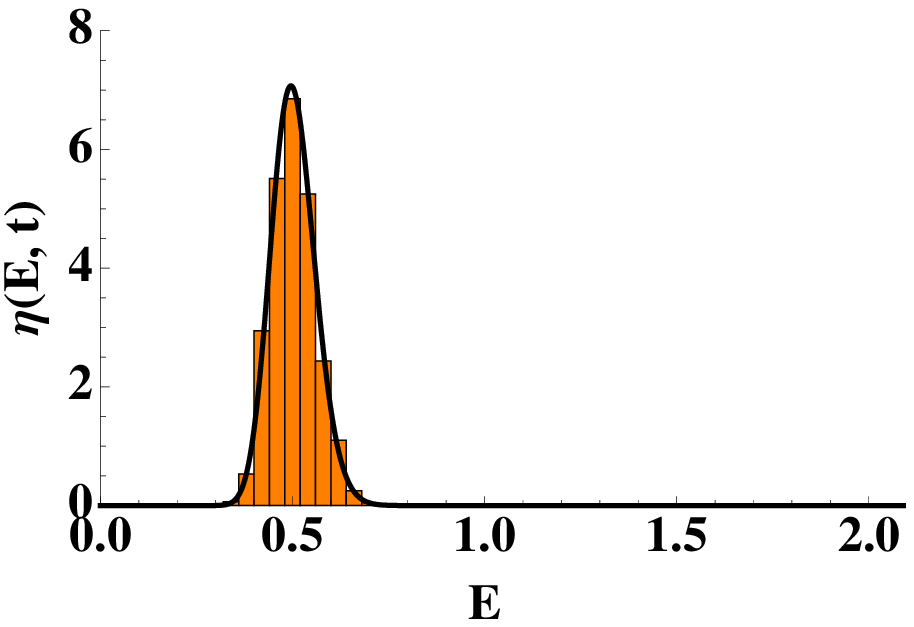}}
\subfloat[$t=1000$]{\label{FUM1D b}\includegraphics[width= 4.0cm] 
{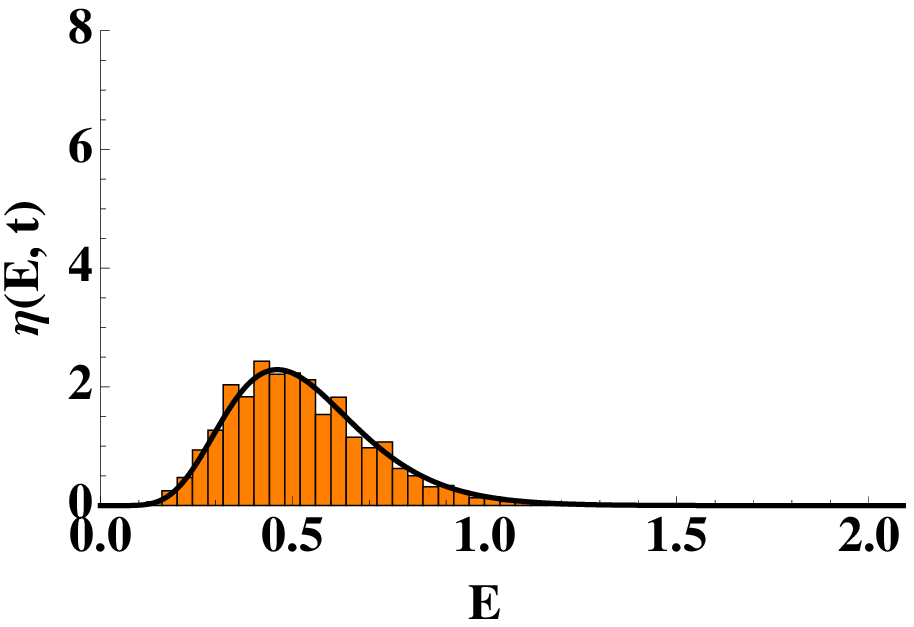}}
\subfloat[$t=5000$]{\label{FUM1D c}\includegraphics[width = 4.0cm] 
{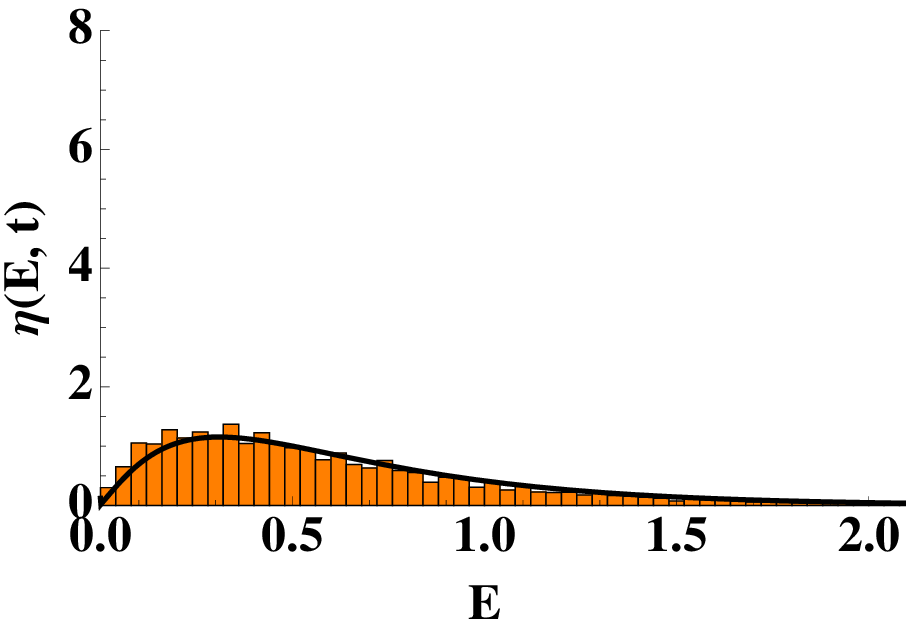}}
\subfloat[$t=15000$] {\label{FUM1D d} \includegraphics [width=4.0cm] 
{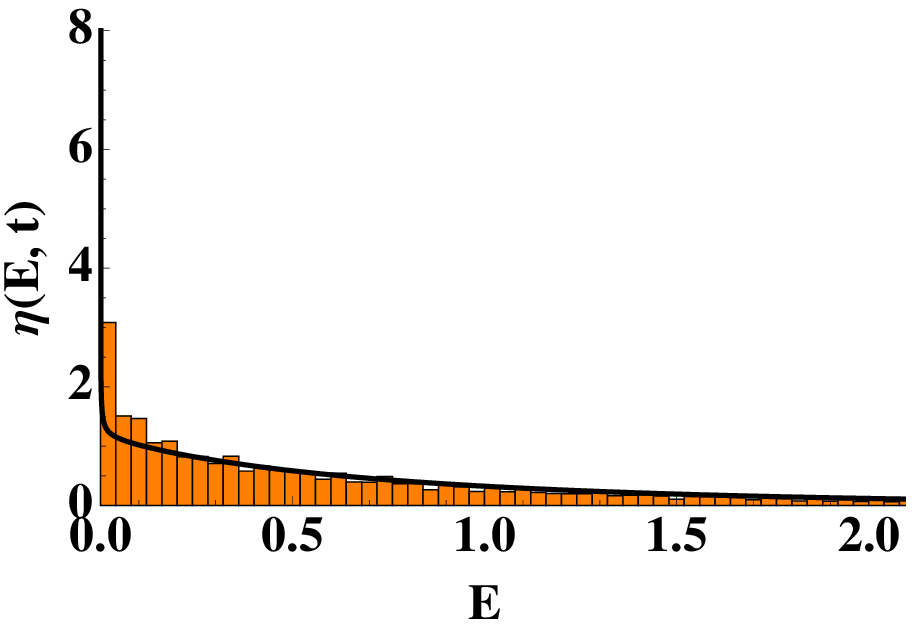}}
\caption{\label{FUM1Ddat} Energy distribution $\eta(E,
t)$ of $10^{5}$ particles following the exact Fermi-Ulam dynamics with small
wall oscillation amplitude $a = 10^{-5}$ at times $t = 100,~ 1000,~ 5000,$ and 
$15000$. The histograms are generated from numerical simulations, and the 
smooth curve is the analytical solution Eq.~(\ref{FPsoln3}) for the energy 
distribution of a particle ensemble in the corresponding quivering billiard.} 
\end{figure}

\begin{figure}[!ht] 
\subfloat[$a=10^{-1}$]{\label{convergePlots a} \includegraphics[width =
4.0cm]{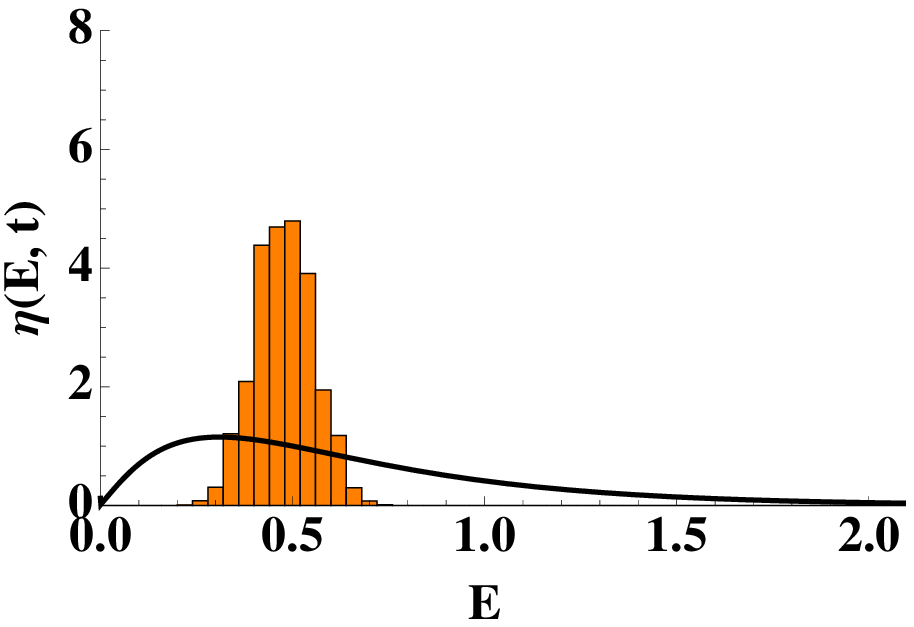}}
\subfloat[$a=10^{-2}$]{\label{convergePlots b} \includegraphics[width = 
4.0cm]{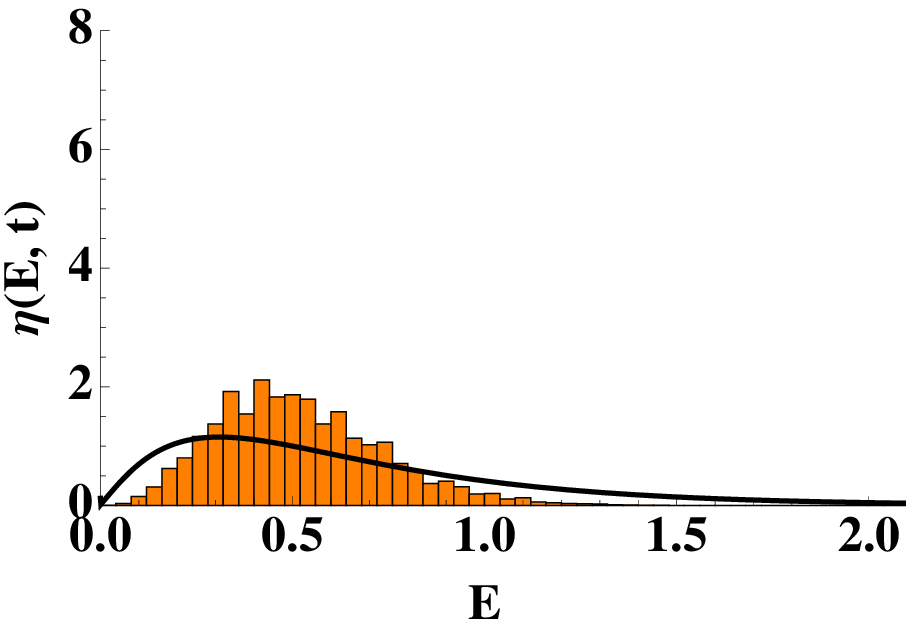}}
\subfloat[$a=10^{-3}$]{\label{convergePlots c}\includegraphics[width = 
4.0cm]{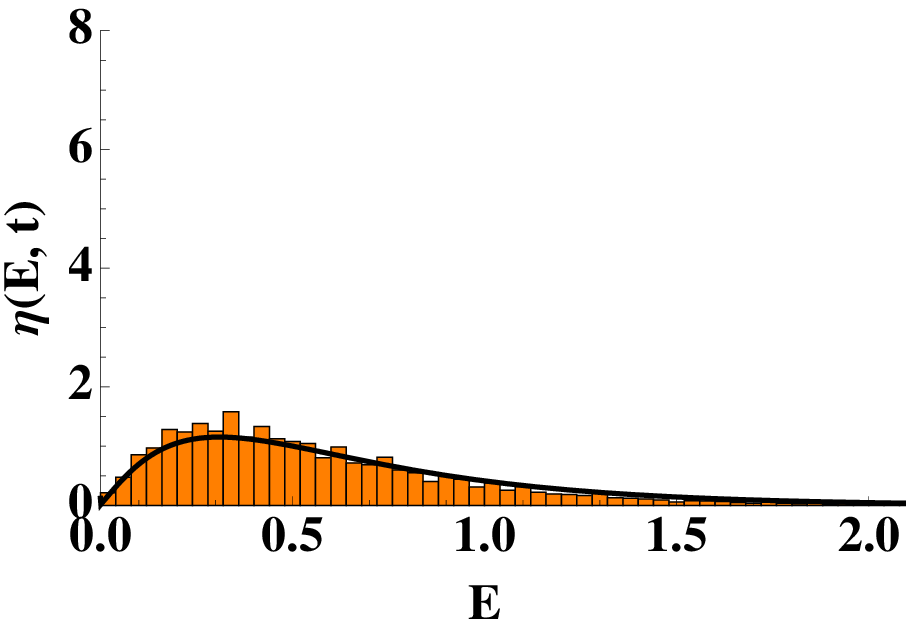}}
\subfloat[$a=10^{-4}$]{\label{convergePlots d}\includegraphics[width =
4.0cm]{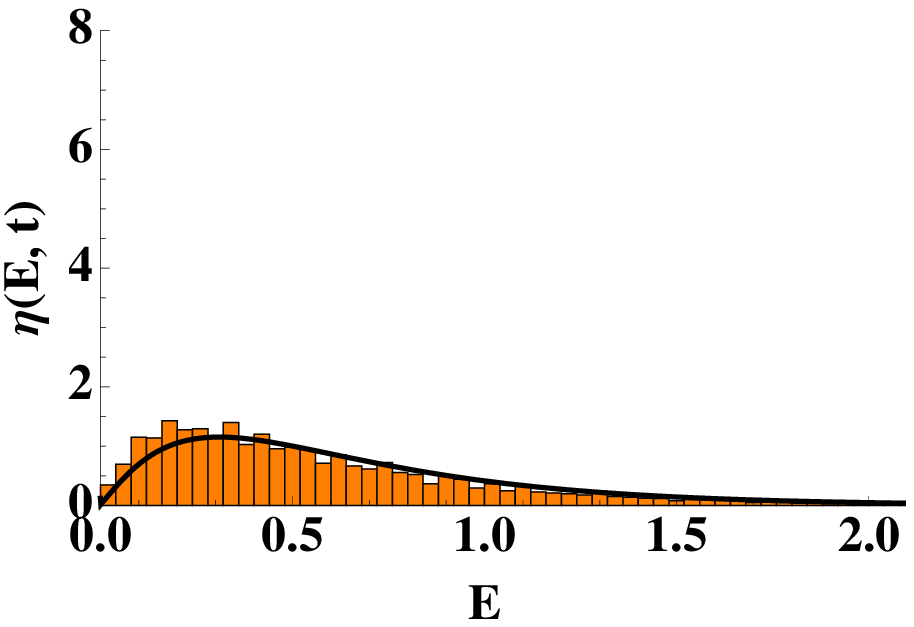}}
\caption{\label{convergePlots} Energy distribution $\eta(E,
t)$ at $t=5000$ of $10^{5}$ particles following the exact Fermi-Ulam dynamics 
for successively smaller wall oscillation amplitudes $a$. The histograms are 
generated from numerical simulations, and the smooth curve is the analytical 
solution Eq.~(\ref{FPsoln3}) for the energy distribution of a particle 
ensemble in the corresponding quivering billiard. The case for $a = 10^{-5}$ 
is shown in the $t=5000$ plot in Fig.~\ref{FUM1Ddat}} 
\end{figure}

\begin{figure}[!h]
\subfloat[$t=100$]{\label{QB2D a} \includegraphics[width = 4.0cm]{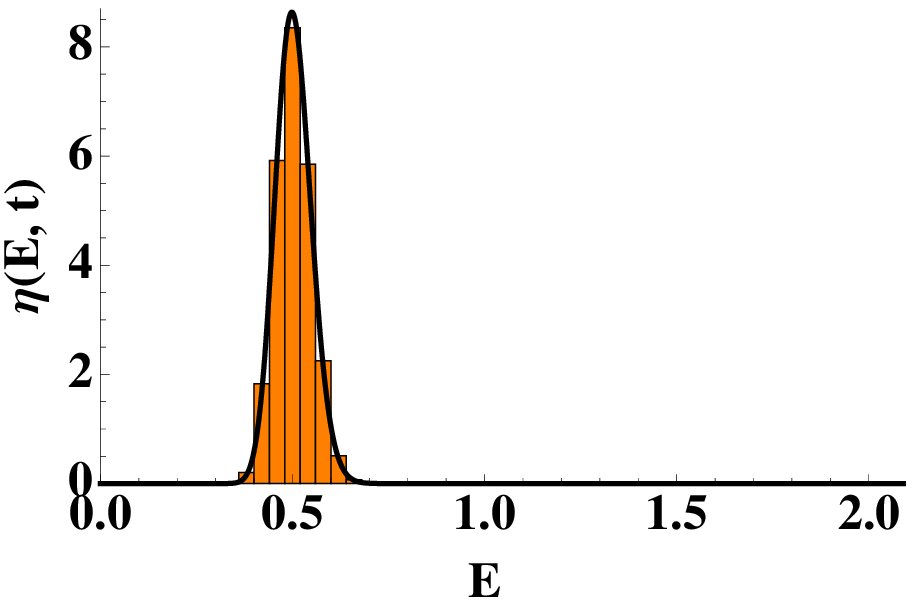}}
\subfloat[$t=1000$]{\label{QB2D b}\includegraphics[width = 4.0cm]{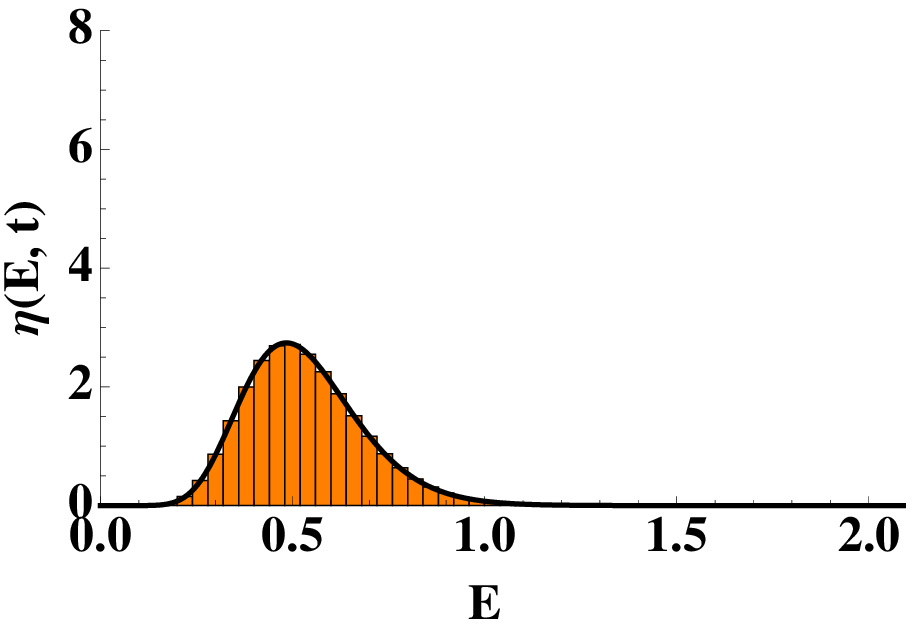}}
\subfloat[$t=5000$]{\label{QB2D c}\includegraphics[width = 4.0cm]{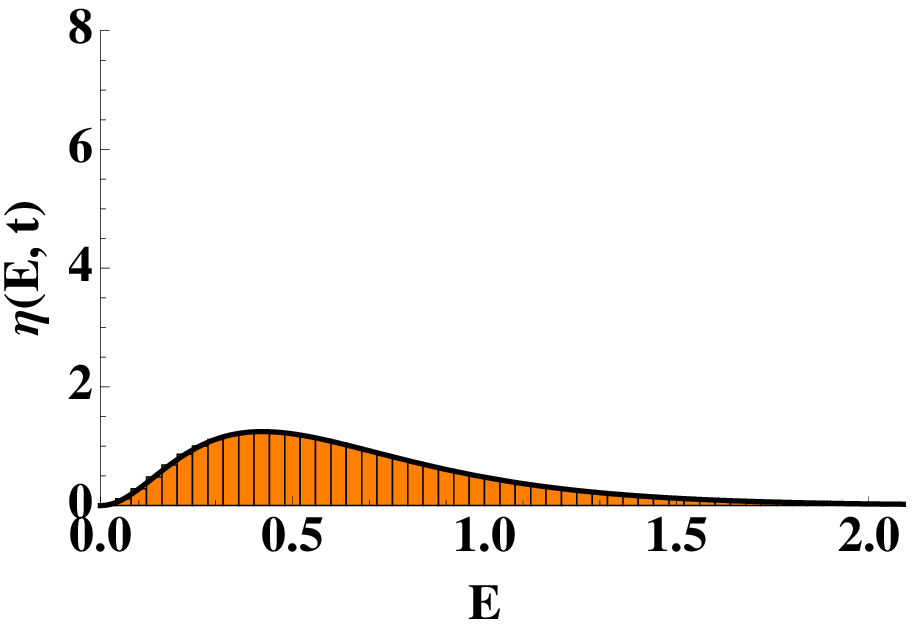}}
\subfloat[$t=15000$]{\label{QB2D d}\includegraphics[width =4.0cm] 
{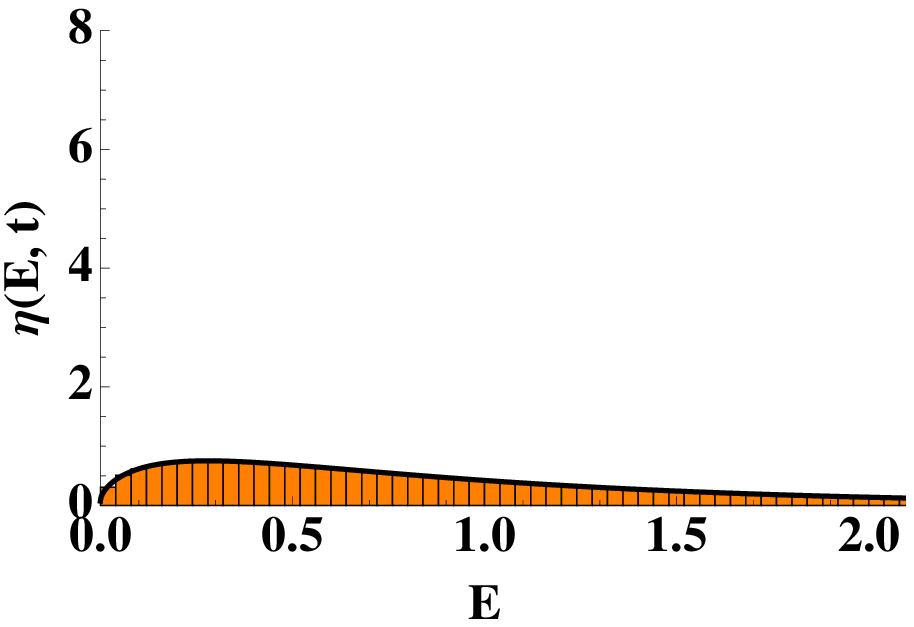}}
\caption{\label{QB2Ddat} Energy distribution $\eta(E, t)$
of $10^{5}$ particles at $t = 100,~ 1000,~ 5000,$ and $15000$ in a
two-dimensional chaotic quivering billiard. The histograms are generated from
numerical simulations, and the smooth curve is the analytical solution for the
energy distribution given by Eq.~(\ref{FPsoln3}).} 
\end{figure}

We now give explicit examples of
quivering billiards in one and two dimensions and support the previous sections'
analyses with numerical work. Consider first a one dimensional Fermi-Ulam model
with one wall oscillating at a constant speed. Following the notation of
Sec.~\ref{sec:II}, the position of the moving wall about its mean position is
given by 
\begin{equation} \label{FUMpos} 
g(t) = 
\begin{cases} 
a[-1 + 4\Psi(t)],& 0 \leq~ \Psi(t) < \frac{1}{2} \\ 
a[1-4(\Psi(t)-\frac{1}{2})], & \frac{1}{2} \leq~ \Psi(t) < 1, 
\end{cases} 
\end{equation}
and the corresponding wall velocity is given by 
\begin{equation} \label{FUMvel} 
u(t) = 
\begin{cases}
4u_{c}, & 0 \leq~ \Psi(t) < \frac{1}{2} \\

-4u_{c}, & \frac{1}{2} \leq~ \Psi(t)< 1. 
\end{cases} 
\end{equation} 
The numerical analyses of this Fermi-Ulam model
are presented in Figs.~\ref{FUM1Ddat} and \ref{convergePlots}. The histograms 
in Fig.~\ref{FUM1Ddat} show of the evolution of the energy distribution of 
$10^{5}$ particles of mass $m = 1$ in a microcanonical ensemble with initial 
speed $v_{0}
= 1$ at time $t=0$, and the curves show the analytical solution for this 
system in the quivering limit as predicted by Eq.~(\ref{FPsoln3}). For this 
simulation, we set $L = 1.0$, $a =10^{-5}$, and $\tau = 10^{-2}$, which gives 
$u_{c} = 10^{-3}$. We see good agreement, with some small deviation apparent 
beginning at $t = 5000$. We suspect that the deviation is due to the faster 
particles interacting with the elliptic islands in phase space, which is not 
accounted for in the quivering
billiard. By the time $t = 15000$, a sufficient number of the particles have
gained enough energy such that the system is no longer approximately quivering.
Further energy gain is stunted by elliptic islands, so we see an excess of
probability (an excess relative to the quivering billiard energy distribution)
begin to build up at low energies. Figure \ref{convergePlots} shows the same
Fermi-Ulam model, with $u_{c} = 10^{-3}$, for successively smaller and smaller
values of $a$ and $\tau$ at time $t=5000$. As $a$ becomes smaller, we see the
actual energy distribution converge to the distribution predicted by the
quivering billiard. 
\\
\begin{figure}[!ht]
\subfloat[$t=100$]{\label{Q1D a}\includegraphics[width = 4.0cm]{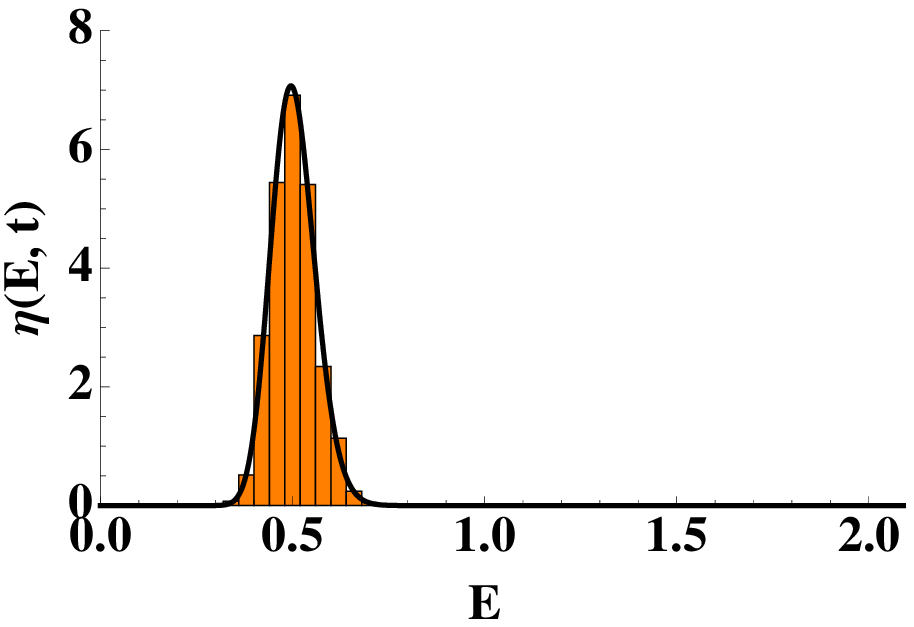}}
\subfloat[$t=1000$]{\label{Q1d b}\includegraphics[width = 4.0cm]{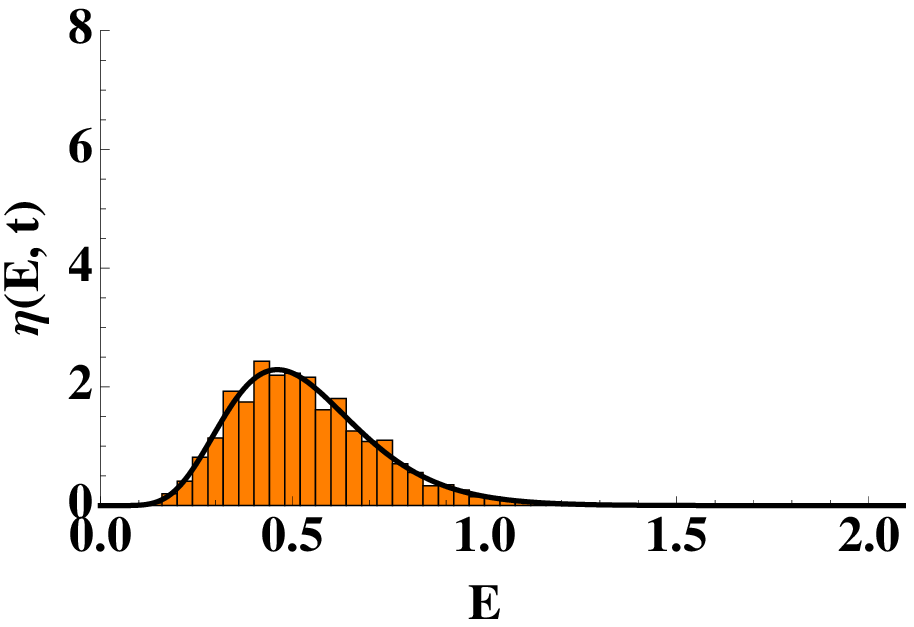}}
\subfloat[$t=5000$]{\label{Q1d c}\includegraphics[width = 4.0cm]{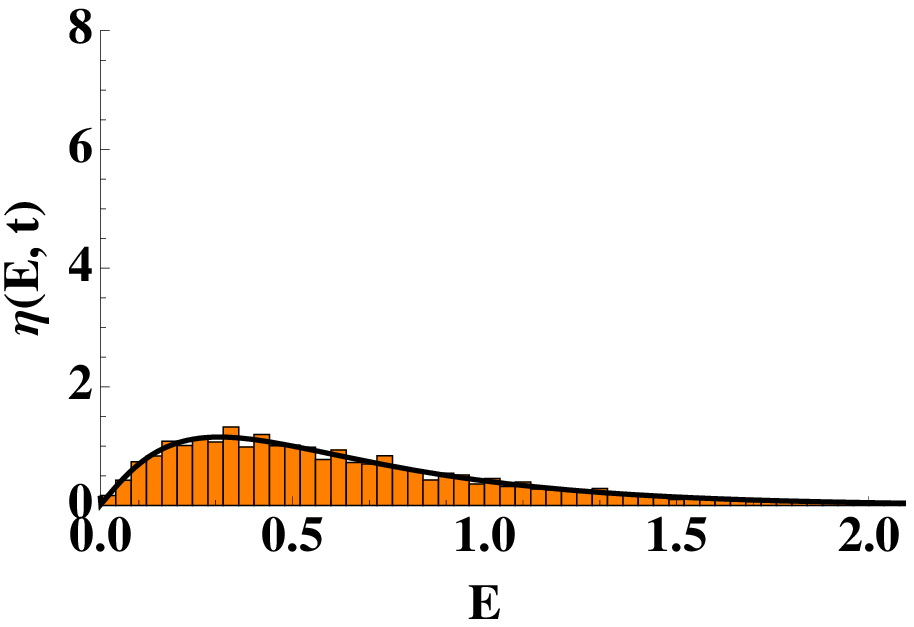}}
\subfloat[$t=15000$]{\label{Q1d d}\includegraphics[width = 4.0cm] 
{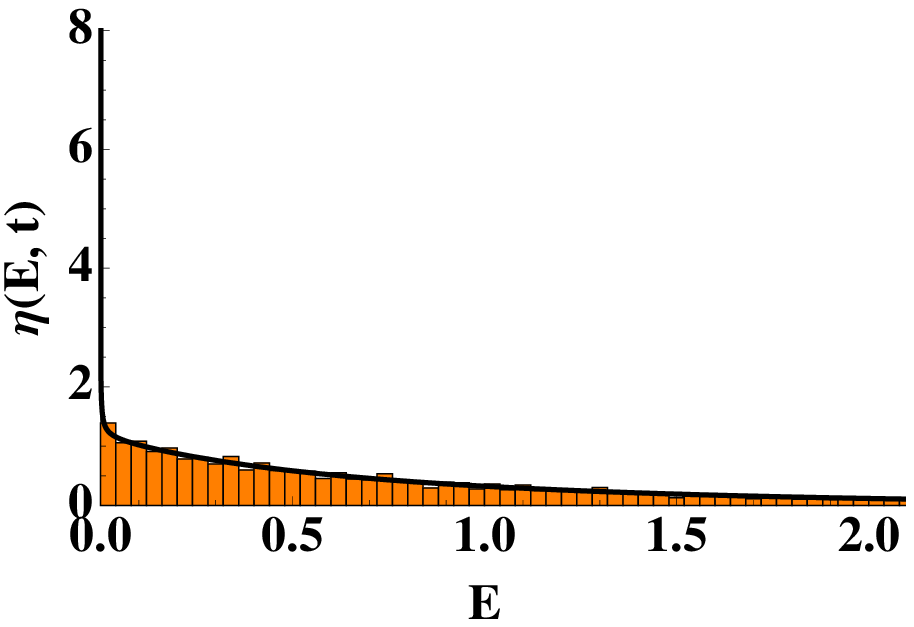}}
\caption{\label{QB1Ddat} Energy
distribution $\eta(E, t)$ of $10^{5}$ particles at $t = 100,~ 1000,~ 5000,$ and
$15000$ in a quivering billiard corresponding to the quivering limit of the
Femi-Ulam model used in Fig.~\ref{FUM1Ddat}. The histograms are generated from
numerical simulations, and the smooth curve is the analytical solution for the
energy distribution given by Eq.~(\ref{FPsoln3}).} 
\end{figure}  
\indent The quivering limit of the Fermi-Ulam model given in 
Eqs.~(\ref{FUMpos}) and (\ref{FUMvel}) is found by following the procedures 
described in Sec.~\ref{sec:II}. We first obtain the unbiased distribution, 
\begin{equation}
P(u_{b}|0) = \frac{1}{2}\delta(u_{b} - 4u_{c}) + \frac{1}{2}\delta(u_{b} +
4u_{c}), 
\end{equation} 
and then the biased distribution $P(u_{b}|v_{b-1})$,
\begin{equation} 
P(u_{b}|v_{b-1}) = 
\begin{cases} \frac{1}{2} \left(1 -
\frac{u_{b}}{v_{b-1}} \right)\left[\delta(u_{b} - 4u_{c}) + \delta(u_{b} +
4u_{c})\right], & v_{b-1} > 4u_{c} \\ 
\delta(u_{b} + 4u_{c}), & v_{b-1} \leq 4u_{c}. 
\end{cases} 
\end{equation} 
The drift and diffusion terms corresponding
to this quivering billiard are found by following the procedures
Sec.~\ref{subsec:IIID}. We note that $M_{2}(\mathbf{q}_{b}) = 16u_{c}^{2}$ for
the moving wall, and $M_{2}(\mathbf{q}_{b}) = 0$ for the stationary wall, so
Eq.~(\ref{coarseVel}) yields $\overline{u^{2}} = (1 / 2) ~ 16u_{c}^{2}$. The
coarse grained free flight distance is given simply by $\overline{l} = L$, so we
find 
\begin{eqnarray} 
g_{1}(E) & = & \frac{32\sqrt{2m} ~ u_{c}^{2}}{L}
E^{\frac{1}{2}} = \alpha E^{\frac{1}{2}} \\ g_{2}(E) & = & \frac{64\sqrt{2m} ~
u_{c}^{2}}{L} E^{\frac{3}{2}} = 2\alpha E^{\frac{3}{2}}. \nonumber
\end{eqnarray} 
The drift and diffusion terms are independent of time, so the
rescaled time $s$ is simply $s = \alpha t$. Using the same values for $L$, $m$,
and $u_{c}$ the we used in the Fermi-Ulam simulation, we find $\alpha \approx
4.53 \times 10^{-5}$. Figure \ref{QB1Ddat} shows the evolving energy
distribution in the simulated quivering billiard, with the analytical result
predicted by Eq.~(\ref{FPsoln3}) superimposed. Our analytical solution agrees
very well with the numerical simulation.
\begin{figure}[!t] 
\includegraphics{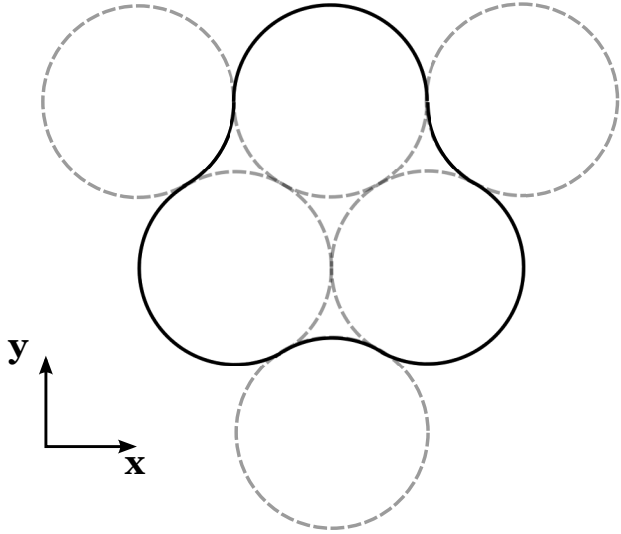} 
\caption{\label{clover}
The six-circle clover billiard, constructed from sections of six adjacent
equi-radii circles.} 
\end{figure} 
\\
\indent For pedagogical purposes, we now construct and simulate a 
two-dimensional quivering billiard. For the billiard shape, we have chosen the 
six-circle clover introduced in Ref.~\cite{Jar1993}, depicted here in 
Fig.~\ref{clover}. We set the normal wall velocities along the billiard boundary to be, 
\begin{equation}
u(\mathbf{q}, t) = 
\begin{cases} u_{c} |\hat{\mathbf{n}}(\mathbf{q}) \cdot
\hat{\mathbf{x}}|, & 0 \leq~ \Psi(t) <\frac{1}{2} \\ 
-u_{c}|\hat{\mathbf{n}}(\mathbf{q}) \cdot \hat{\mathbf{x}}|, & \frac{1}{2} \leq~
\Psi(t) < 1, 
\end{cases}
\end{equation}
where $\hat{\mathbf{n}}(\mathbf{q})$ is
the outward unit normal to the wall at $\mathbf{q}$ and $\hat{\mathbf{x}}$ is
the unit vector in the x-direction. This choice of wall velocities gives in the
quivering limit, 
\begin{equation} 
P(u_{b}|0,\mathbf{q}_{b}) =
\frac{1}{2}\delta\boldsymbol{(}u_{b} - u_{c} |\hat{\mathbf{n}}(\mathbf{q}_{b})
\cdot \hat{\mathbf{x}}|\boldsymbol{)} + \frac{1}{2}\delta\boldsymbol{(}u_{b} +
u_{c} |\hat{\mathbf{n}}(\mathbf{q}_{b}) \cdot \hat{\mathbf{x}}|\boldsymbol{)}
\end{equation} 
\begin{equation} 
P(u_{b}|v_{b-1},\mathbf{q}_{b},\theta_{b}) =
\begin{cases} \left(1 - \frac{u_{b}}{v_{b-1 \sin(\theta_{b})}} \right)
P(u_{b}|0,\mathbf{q}_{b}), & v_{b-1} > u_{c}|\hat{\mathbf{n}}(\mathbf{q}_{b})
\cdot \hat{\mathbf{x}}| \\ 
\delta\boldsymbol{(}u_{b} + u_{c}
|\hat{\mathbf{n}}(\mathbf{q}_{b}) \cdot \hat{\mathbf{x}}|\boldsymbol{)}, &
v_{b-1} \leq u_{c}|\hat{\mathbf{n}}(\mathbf{q}_{b}) \cdot \hat{\mathbf{x}}|.
\end{cases} 
\end{equation} 
The six-circle clover constructed from equi-radii
circles is fully chaotic \cite{Jar1993}, so over time scales greater than the
clover's ergodic time scale, $\overline{u^{2}}$ is just $M_{2}(\mathbf{q})$
averaged uniformly over the billiard boundary. For any $\mathbf{q}$ on the
boundary, we have $M_{2}(\mathbf{q}) = u_{c}^{2}~ |\hat{\mathbf{n}}(\mathbf{q})
\cdot \hat{\mathbf{x}}|^{2}$, and from Fig.~\ref{clover}, we see the outward
normals $\hat{\mathbf{n}}(\mathbf{q})$ are distributed uniformly around a unit
circle, so we have $\overline{u^{2}} = (1 / 2) ~u_{c}^{2}$. The coarse grained
free flight distance $\overline{l}$, over time scales greater than the ergodic
time scale, is just the billiard's mean free path. For a two dimensional ergodic
billiard, the mean free path is given by $ \pi A / S$, where $A$ is the
billiard's area and $S$ is the billiard's perimeter \cite{Jar1993}. If the
radius of the circles used to construct the six-circle clover is $R$, then the
geometry of Fig.~\ref{clover} gives $A = R^{2}(4 \sqrt{3} + \pi)$ and $S = 4 \pi
R$. We thus have for the drift and diffusion coefficients, 
\begin{eqnarray}
g_{1}(E) & = & \frac{2\sqrt{2m} ~ u_{c}^{2}}{l} E^{\frac{1}{2}} = \alpha
E^{\frac{1}{2}} \\ 
g_{2}(E) & = & \frac{8\sqrt{2m} ~ u_{c}^{2}}{3~l}
E^{\frac{3}{2}} = \frac{4}{3}\alpha E^{\frac{3}{2}}. \nonumber 
\end{eqnarray}
where $l = R(\sqrt{3} + \pi / 4)$ is the mean free path. 
\\ 
\indent
Figure~\ref{QB2Ddat} shows the energy evolution of a microcanonical ensemble of
$10^{5}$ particles in a quivering clover, with the distribution
Eq.~(\ref{FPsoln3}) superimposed. The particles have mass $m=1$ and initial
energy $E_{0} = 1 / 2$. We constructed the clover with circles of radius $R = 1$
and set $u_{c} = 6.35 \times 10^{-3}$ to give $\alpha \approx 4.53 \times
10^{-5}$. Again, we see good agreement between the distribution predicted by
Eq.~(\ref{FPsoln3}) and the simulated energy distribution.

\section{Summary and Conclusions} \label{sec:VI} 
In this work, we have defined a
particular fixed wall limit of time-dependent billiards, the quivering limit,
and explored the evolution of particles and ensembles in the resulting quivering
billiards. We have conjectured that any physically consistent, non-trivial, 
fixed wall 
limit of a time-dependent
billiard must be physically equivalent to the quivering limit, and we have 
shown that the simplifications allowed by a physically
consistent fixed wall limit come at a price: deterministic billiard dynamics
become inherently stochastic. Although quivering is an idealized limit of
billiard motion, we have shown that for smaller and smaller oscillation
amplitudes and periods, time-dependent billiards become better and better
approximated by quivering billiards. Billiards that quiver or approximately
quiver behave universally; particle energy evolves diffusively, particle
ensembles achieve a universal asymptotic energy distribution, and quadratic
Fermi acceleration always occurs, regardless of a billiard's dimensionality or
frozen dynamics. The mechanism for this quadratic Fermi acceleration is
analogous to a resistive friction-like force, present due to the fluctuations
induced by the erratic wall motion, as described by the fluctuation-dissipation
relation in Eq.~(\ref{FlucDiss1}). 
\\ 
\indent Through this work, we have gained
some insight into issues that have been discussed in the previous literature.
Namely, we concluded that in the quivering limit, the quasilinear approximation
is exact, not an approximation. Also, we showed that the often used static wall
approximation fails because it is unphysical and can not take into account the
statistical bias towards inward moving wall collisions. Energy gain in the
static wall approximation is a purely mixing effect; unbiased fluctuations in
particle velocity produce an average increase in particle velocity squared,
analogous to a Brownian random walk where unbiased fluctuations in position
produce an average increase in squared distance from the initial position. From
this observation, and the fact that the static wall approximation gives one half
the asymptotic energy growth rate observed in exact systems, we conclude that in
the quivering limit, half of the average energy gain observed in a
time-dependent billiard is due to the mere presence of fluctuations, and half is
due to the fact that energy gaining fluctuations are more likely than energy
losing fluctuations. 
\\ 
\indent We close by acknowledging that we have not given
a rigorous mathematical proof showing that deterministic time-dependent
billiards become stochastic quivering billiards in the quivering limit. One
possible approach toward such a proof would be to define some sort of space of
time-dependent billiards consisting of systems with different oscillation
amplitudes and periods, define a metric to give some notion of distance in this
space, and prove that particular sequences in this space with successively
smaller amplitudes and periods are Cauchy sequences. One could then determine
what properties the space of systems would need to posses in order to assure
that these Cauchy sequences converge to limits, and then study the limits by
studying the sequences that converge to them. Instead of a rigorous mathematical
approach, we have taken a more intuitive approach and have attempted to justify
our work by using physical reasoning and by showing consistency with previous
results. We hope that the evidence is convincing enough to mitigate our
mathematical deficiencies.

\begin{acknowledgments} The authors would like to thank Zhiyue Lu and Kushal
Shah for useful discussions. This work was supported by the U. S. Army 
Research Office under contract number W911NF-13-1-0390.
\end{acknowledgments}

\clearpage \appendix* \section{} 
\begin{figure}[!htb]
\includegraphics{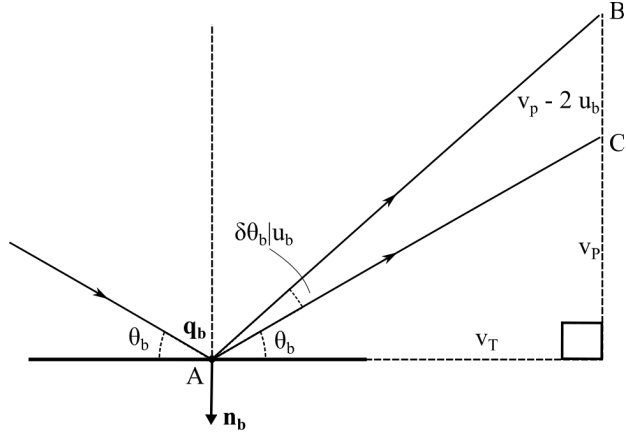}
\caption{\label{CollisionVelocitySpace}Incoming and outgoing particle
trajectories at the $b^{th}$ collision location $\mathbf{q}_{b}$ in the full
and frozen dynamics, assuming a collision wall velocity $u_{b}$. The full
dynamics trajectory is perturbed by an angle $\delta \theta|u_{b}$ relative to
the frozen dynamics trajectory. $\mathbf{n}_{b}$ is the outward normal to the
boundary at $\mathbf{q}_{b}$.} 
\end{figure} 
\begin{figure}[!htb]
\includegraphics[width = 8.0cm]{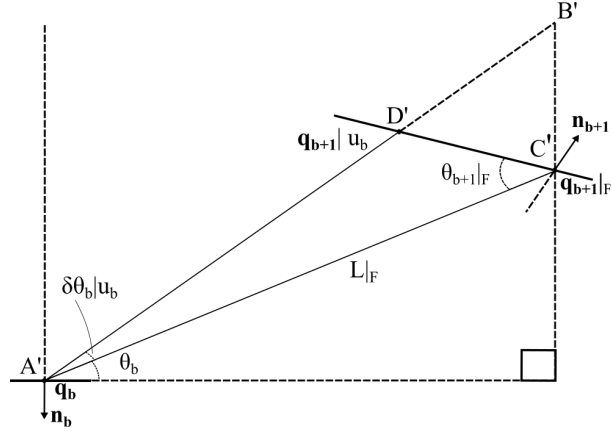}
\caption{\label{CollisionGeometry} The geometrical relationship between
$\mathbf{q}_{b}$, $\mathbf{q}_{b+1}|_{F}$, and $\mathbf{q}_{b+1}|u_{b}$.
$\mathbf{q}_{b}$ and $\mathbf{q}_{b+1}|_{F}$ denote the $b^{th}$ and
$(b+1)^{th}$ collision locations in the frozen dynamics, respectively, while
$\mathbf{q}_{b+1}|u_{b}$ denotes the $(b+1)^{th}$ collision location in the full
dynamics. $\mathbf{n}_{b}$ and $\mathbf{n}_{b+1}$ are the outward normals to
the boundary at $\mathbf{q}_{b}$ and $\mathbf{q}_{b+1}|_{F}$, respectively.}
\end{figure} 
Here, we find $\|\delta\mathbf{q}_{b+1}|u_{b}\|$, the magnitude of
the perturbation to the frozen dynamics $(b+1)^{th}$ collision location due to 
the energy gained or lost at the $b^{th}$ collision in the full dynamics. In the
frozen dynamics, the collision angle $\theta_{b}$ is equal to the angle of
reflection. Let $\theta_{b} + \delta \theta|u_{b}$ be the reflected angle in the
full dynamics, assuming a wall velocity of $u_{b}$ at the $b^{th}$ collision. We
denote $v_{b-1}$ as the incoming particle speed at the $b^{th}$ collision,
$v_{T}$ as the velocity component tangent to the wall, $v_{P}$ as the reflected
particle's velocity component perpendicular to the wall in the frozen dynamics,
and $v_{P}|u_{b}$ as the reflected perpendicular velocity component in the full
dynamics. The collision kinematics give $v_{P}|u_{b} = v_{P} - 2 u_{b}$. The
perturbation $\delta \theta|u_{b}$ can be found using the geometry in
Fig.~\ref{CollisionVelocitySpace}. Note that $\tan(\theta_{b}) =
\frac{v_{P}}{v_{T}}$ and $\tan(\theta_{b} + \delta\theta|u_{b}) =
\frac{v_{P}|u_{b}} {v_{T}}$. Expanding $\tan(\theta_{b} + \delta\theta|u_{b})$
to first order in $\delta\theta|u_{b}$, we find
\begin{eqnarray}
\tan(\theta_{b} + \delta\theta|u_{b}) & = & \frac{v_{P}|u_{b}} {v_{T}} \\ 
 & = & \tan(\theta_{b}) + \frac{1}{\cos^{2}(\theta_{b})} \delta \theta|u_{b}
 \nonumber \\ 
 & = & \frac{v_{P}}{v_{T}} + \frac{1}{\cos^{2}(\theta_{b})}\delta
\theta|u_{b}. \nonumber 
\end{eqnarray} 
Noting that $v_{P}|u_{b} = v_{P} - 2
u_{b}$ and $v_{T} = v_{b-1} \cos{\theta_{b}}$, we solve for $\delta
\theta|u_{b}$ to find
\begin{equation} 
\delta \theta|u_{b} = 2 \cos(\theta_{b}) \frac{u_{b}}{v_{b-1}}.
\end{equation}
\indent Figure~\ref{CollisionGeometry}
shows the geometry of the $b^{th}$ and $(b+1)^{th}$ collisions in both the full
and frozen dynamics, where $\|\delta\mathbf{q}_{b+1}|u_{b}\|$ is the length of
the line segment $C'D'$. We assume that $\delta \theta|u_{b}$ is small enough
such that the wall appears flat between the frozen and full dynamics' 
$(b+1)^{th}$ collision locations. The triangle $A'B'C'$ in 
Fig.~\ref{CollisionGeometry} is similar to the triangle $ABC$ in 
Fig.~\ref{CollisionVelocitySpace}, so we have 
$\frac{|BC|}{|AC|} = \frac{|B'C'|}{|A'C'|} = \frac{2|u_{b}|}{v_{b-1}}$. We note
that $|A'C'|$ is the distance between the $b^{th}$ and $(b+1)^{th}$ collision
locations in the frozen dynamics, so we denote $|A'C'| = L_{b}|_{F}$ and find
\begin{equation} 
|B'C'| = \frac{2 |u_{b}|}{v_{b-1}} L_{b}|_{F}. 
\end{equation}
All angles in Fig.~\ref{CollisionVelocitySpace} can be found in terms of
$\theta_{b}$, $\theta_{b+1}|_{F}$, and $\delta \theta|u_{b}$. By applying the
Law of Sines to the triangle $B'C'D'$, we find
\begin{equation} 
|C'D'| = 2
L_{b}|_{F} \frac{\cos(\theta_{b})}{\sin(\theta_{b+1}|_{F})}
\frac{|u_{b}|}{v_{b-1}}.
\end{equation} 
We thus have 
\begin{equation}
\|\delta\mathbf{q}_{b+1}|u_{b}\| = 2 L_{b}|_{F}
\frac{\cos(\theta_{b})}{\sin(\theta_{b+1}|_{F})} \frac{|u_{b}|}{v_{b-1}}.
\end{equation}

\bibliographystyle{apsrev4-1}
\bibliography{QuiveringBilliard-9-14-2015}

\end{document}